\documentclass[prl, aps, groupedaddress,superscriptaddress,notitlepage, twocolumn]{revtex4-1}
\usepackage[utf8x]{inputenc}
\usepackage{color}
\usepackage{bbm} 
\usepackage{nicefrac}
\usepackage{amsfonts,amsmath,amssymb,stmaryrd}
\usepackage{braket}
\usepackage[autostyle=true]{csquotes}
\usepackage{tabularx}
\usepackage{multirow} 
\usepackage{hhline}
\usepackage{graphicx}
\usepackage{subfigure}  %
\usepackage{bbm} 
\usepackage{hyperref}
\usepackage[pdftex]{epsfig}
\usepackage{mathrsfs}
\usepackage{verbatim}
\usepackage{centernot}
\usepackage{ulem}
\usepackage{array}
\usepackage{cancel}
\usepackage{ifthen}
\usepackage{bm}	%
\usepackage{siunitx}
\usepackage{todonotes}
\InputIfFileExists{gitinfo-latex.inc}{}{}
\usepackage{bbold}
\usepackage{listings}
\usepackage{color}
\usepackage[acronym,nomain]{glossaries}

\newcolumntype{L}{>{$}l<{$}} %
\newcolumntype{C}{>{$}c<{$}} %

\newcommand{\TRb}{T_{\mathrm{Rb}}}
\newcommand{\TCs}{T_{\mathrm{Cs}}}

\newcommand*\diff{\mathop{}\!\mathrm{d}}

\newcommand{\Hint}{\hat{H}^{\text{int}}}
\newcommand{\hint}{\hat{H}^{\mathrm{int}}}

\newcommand{\mF}{{m_F}}
\newcommand{\mFCs}{{m_{F, \mathrm{Cs}}}}
\newcommand{\FCs}{F_{\mathrm{Cs}}}
\newcommand{\FRb}{F_{\mathrm{Rb}}}
\newcommand{\mFRb}{{m_{F, \mathrm{Rb}}}}

\newcommand{\PF}[1]{\mathcal{P}_{ #1 }}

\newcommand{\kB}{k_B}
\newcommand{\mCs}{m_{\mathrm{Cs}}}
\newcommand{\mRb}{m_{\mathrm{Rb}}}

\newcommand{\Gammael}{{\Gamma_{\mathrm{el}}}}
\newcommand{\Gammase}{{\Gamma_{\mathrm{se}}}}
\newcommand{\spinrate}{\Gamma}
\newcommand{\lossrate}{\Lambda}

\newcommand{\densrb}{n_\text{Rb}}

\newcommand{\denscs}{n_\text{Cs}}

\newcommand{\mf}{{m_F}}

\newcommand{\nexp}{\left<n\right>}

\newcommand{\tukl}{Department of Physics and Research Center OPTIMAS, Technische Universit\"at Kaiserslautern, 67663 Kaiserslautern, Germany}
\newcommand{\unihannover}{Institut für Quantenoptik, Leibniz Universit\"at Hannover, 30167 Hannover, Germany}

\newcommand{\FigSone}{
	\begin{figure}[t]
		\centering
		\includegraphics[width=.33 \textwidth]{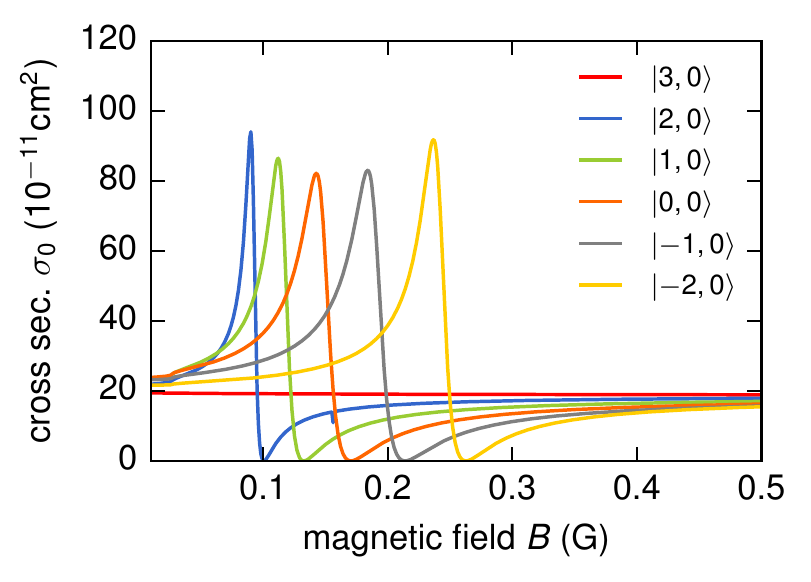}
		\includegraphics[width=.33 \textwidth]{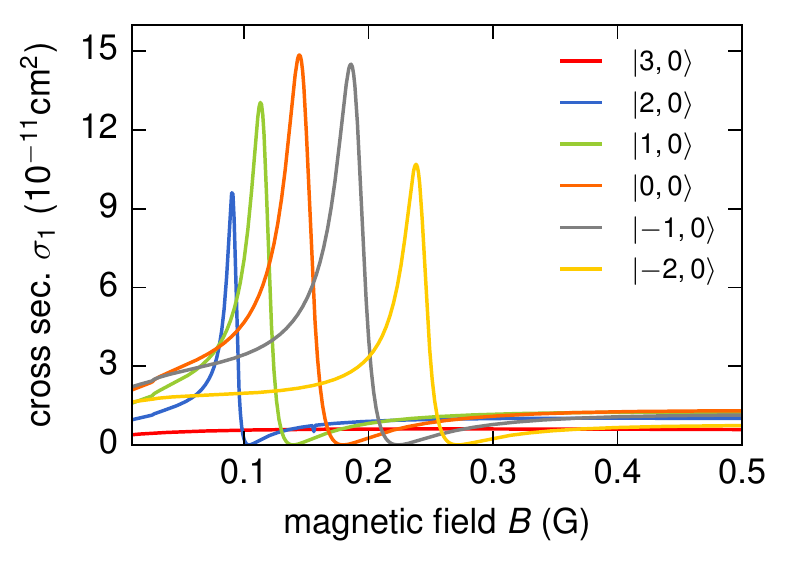}
		\caption{Scattering cross section of elastic collisions (top), spin-exchange of $1\hbar$ (bottom) for Rb in the $m_F=0$ state.
			Data calculated for fixed collision energies $\kB \times \SI{450}{\nano \kelvin}$.}
		\label{fig:RatesNumeric_mF0}
	\end{figure}
}

\newcommand{\FigStwo}{
	\begin{figure}[tb]
		\centering
		\includegraphics[width=.33\textwidth]{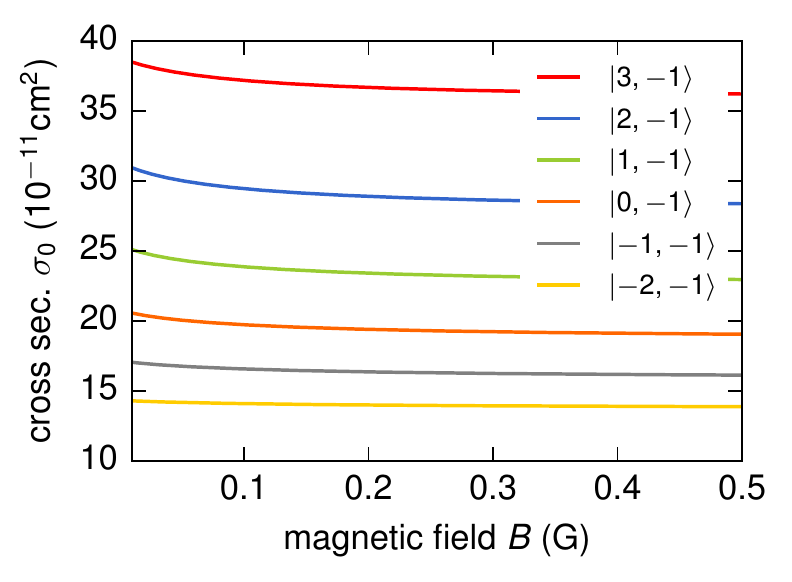}
		\includegraphics[width=.33\textwidth]{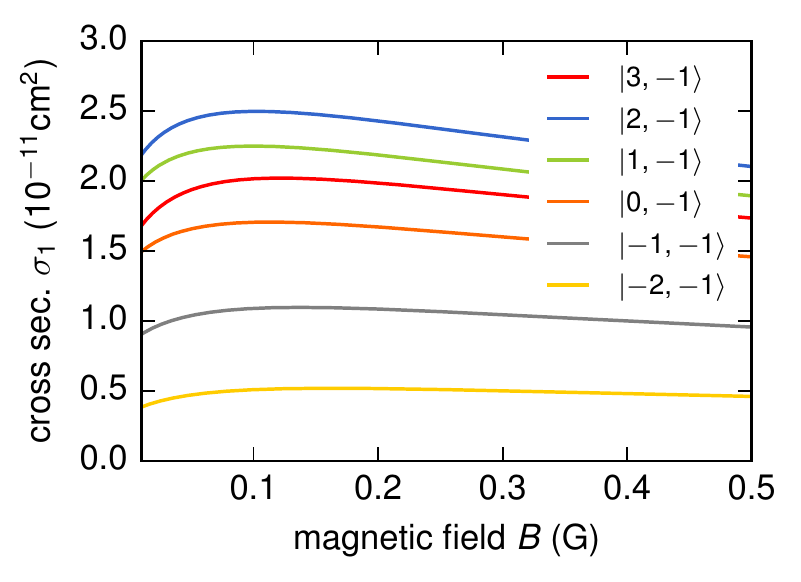}
		\includegraphics[width=.33\textwidth]{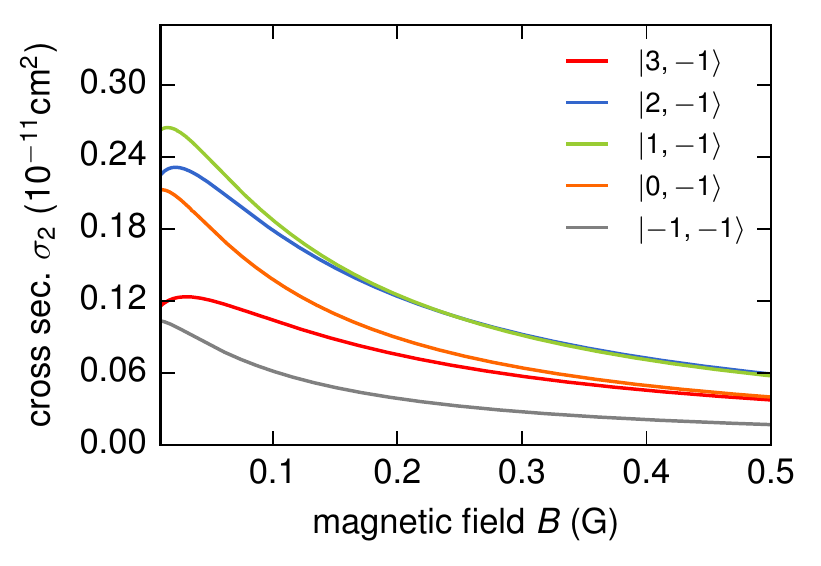}
		\caption{Scattering cross section for spin-exchange of elastic (top), $1\hbar$ (middle), and $2\hbar$ (bottom) for Rb in the $m_F=-1$ state.
			Data calculated for fixed collision energies $\kB \times \SI{450}{\nano \kelvin}$.}
		\label{fig:RatesNumeric_mFm1}
	\end{figure}
}

\newcommand{\FigSthree}{
	\begin{figure}[h]
		\centering
		\includegraphics[scale=1]{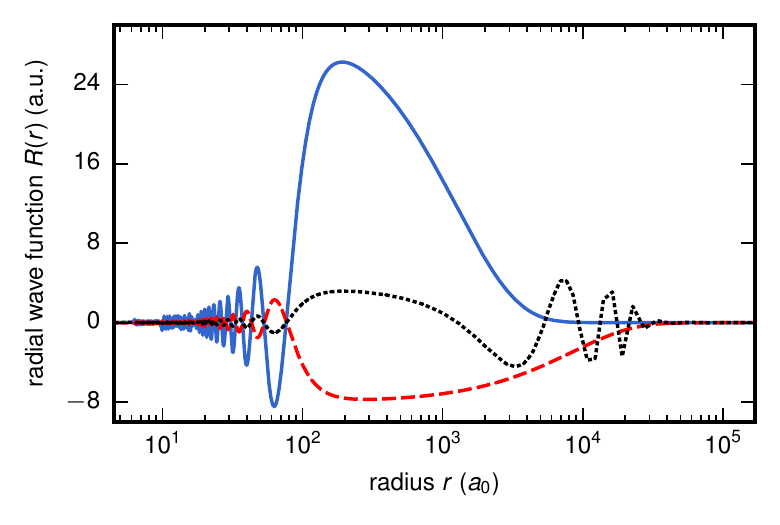}
		\caption{Radial wavefunction $R$ contributions to the $n=-1$ state of the Cs-Rb triplet potential state with total angular momentum $\ket{F=2, M=2}$ for $B=\SI{80}{\milli G}$.
			Three asymptotic states mainly contribute, which are $\ket{\FCs=3, \mFCs=3} + \ket{\FRb=1, \mFRb=-1}$ (blue, solid), $\ket{\FCs=3, \mFCs=2} + \ket{\FRb=1, \mFRb=0}$ (red, dashed), and $\ket{\FCs=3, \mFCs=1} + \ket{\FRb=1, \mFRb=1}$ (\changed{black}, dotted).
		}
		\label{fig:feshbachStates}
	}
	
\newcommand{\FigSfour}{
	\begin{figure*}[h]
		\centering
		\includegraphics[scale=1]{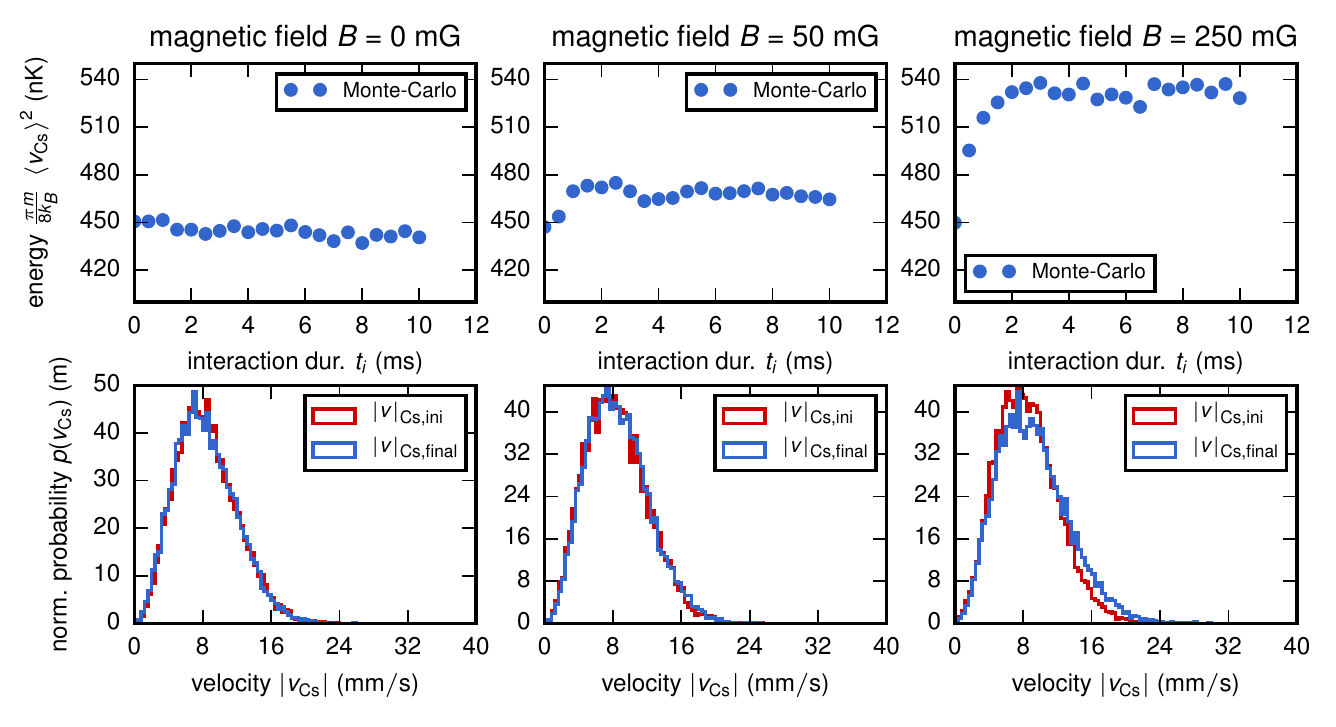}
		\caption{Influence of SE on Cs temperature.
			Cs atoms interact with the Rb bath ($\TRb=\SI{450}{\nano \kelvin}$, homogeneous density $n=\SI{2.5e13}{\centi \meter^{-3}}$) via elastic collisions and SE. 
			SE collisions release energy into the system of $Q = (g_{F, \mathrm{Cs}} - g_{F, \mathrm{Rb}}) \mu_B B = h \times \SI{350}{\kilo \hertz \per G} \times B$ in each SE event.
			(top row) Averaged Cs energy for various interaction durations $t$ at three different magnetic fields.
			SE increases the average energy and the energy equilibrates at a new steady state.
			Typical timescales in the system are the inverse rates of elastic and spin-exchange collisions, $\tau_{\mathrm{el}} = \SI{0.18}{\milli \second}$ and $\tau_{\mathrm{se}} = \SI{3.1}{\milli \second}$, respectively.
			(bottom row) Velocity distribution of initial and final Cs states. 
		}
		\label{fig:thermalization}
	\end{figure*}
}

\newcommand{\methods}{\cite{Supplementary}}
\newcommand{\changed}[1]{\textcolor{black}{#1}}
\renewcommand{\sout}[1]{\unskip}

\begin{document}
	
	\title{Tailored single-atom collisions at ultra-low energies}
	
	\author{Felix Schmidt}
	\affiliation{\tukl}
	
	\author{Daniel Mayer}
	\affiliation{\tukl}
	
	\author{Quentin Bouton}
	\affiliation{\tukl}
	
	\author{Daniel Adam}
	\affiliation{\tukl}
	
	\author{Tobias Lausch}
	\affiliation{\tukl}
	
	\author{Jens Nettersheim}
	\affiliation{\tukl}
	
	\author{Eberhard Tiemann}
	\affiliation{\unihannover}
	
	\author{Artur Widera}
	\email{email: widera@physik.uni-kl.de}
	\affiliation{\tukl}
	\affiliation{Graduate School Materials Science in Mainz, Gottlieb-Daimler-Strasse 47, 67663 Kaiserslautern, Germany}

\date{\today}
	\date{\today}
\begin{abstract}

	We employ collisions of individual atomic Cesium (Cs) impurities with an ultracold Rubidium (Rb) gas to probe atomic interaction with hyperfine- and Zeeman-state sensitivity.
Controlling the Rb bath's internal state yields access to novel phenomena observed in inter-atomic spin-exchange. 
These can be tailored at ultra-low energies, owing to the excellent experimental control over all relevant energy scales.	
First, detecting spin-exchange dynamics in the Cs hyperfine state manifold, we resolve a series of previously unreported Feshbach resonances at magnetic fields below $300\,$mG, separated by energies as low as $h\times \SI{15}{\kilo \hertz}$. 
The series originates from a coupling to molecular states with binding energies below  $h\times \SI{1}{\kilo \hertz}$ and wave function extensions in the $\SI{}{\micro \meter}$ range. 
Second, at magnetic fields below  $\approx 100\,$mG, we observe the emergence of a new reaction path for alkali atoms, where in a single, direct collision between two atoms two quanta of angular momentum can be transferred. 
This path originates from the hyperfine-analogue of dipolar spin-spin relaxation.
Our work yields control of subtle ultra-low-energy features of atomic collision dynamics, opening new routes for advanced state-to-state chemistry, for controlling spin-exchange in quantum many-body systems for solid state simulations, or for determination of high-precision molecular potentials.
\end{abstract}	
\maketitle

Understanding and controlling collisions of two atoms at ultra-low energies are the basis of quantum engineering \cite{Moses2016, Bohn2017}, chemistry \cite{Baranov2012, Quemener2012} and metrology \cite{Safronova2018} applications.	
Advances in cooling and trapping of atoms have opened experimental routes to study atomic interactions with well-defined quantum states at ultracold temperatures \cite{Weiner1999}. 
The energy scale and resolution of individual collision and reaction processes are set by the thermal broadening in a finite-temperature system, as well as the system's lifetime.
Collisional spectroscopy involving single ions \cite{Hall2012, Wolf2017, Sikorsky2018} or atoms \cite{Liu2018} are capable of tracing individual collision pathways or spin-controlled collisions, with an associated energy resolution in the order of few $h \times \si{\giga \hertz}$.
Recently, also Rydberg excitations in cold gases yield access to single ion-atom collisions \cite{Schlagmuller2016, Kleinbach2018}, where an energy resolution in the $h \times \si{\mega \hertz}$ regime is achieved for excitation lifetimes of few $\si{\micro \second}$.
By contrast, collisional energies for neutral atom mixtures at ultra-low temperatures are in the order of few $h \times \si{\kilo \hertz}$.
These low collisional energies have been employed, e.g. to determine scattering phase shifts in an atomic clock \cite{Bennett2017} or to build ultracold molecules from pairs of atoms in optical lattices \cite{Bohn2017, Danzl2010}. 
Hence, probing the interspecies interaction of individual collisional channels should be possible with unprecedented resolution, where we focus on scattering processes of unbound ultracold atoms in close proximity to the dissociation threshold, rather than bound molecular states \cite{Ospelkaus2010, Molony2014}.

Low energies and internal state resolution yield access to intriguing phenomena of a single atom-atom collision that originate from the complex interplay of collisional, Zeeman, hyperfine and molecular interaction energies.
Particularly, we reveal ultra-low-energy features of inter-species spin dynamics, interfacing an ultracold $^{87}\mathrm{Rb}$ bath with single neutral $^{133}\mathrm{Cs}$ atoms \cite{Mayer2018, Schmidt2018} via $s$-wave collisions.
A hierarchy of rates for different collisional processes is identified, i.e. elastic collisions and spin-exchange (SE) in quanta of $1\hbar$ and $2\hbar$, where respective cross sections $\sigma$ roughly scale as $\sigma_{\mathrm{el}} \approx 10 \sigma_{1} \approx 10^2 \sigma_{2}$. 
We adjust both the internal Rb state and the magnetic field value to address specific regimes of SE.
\begin{figure*}
	\begin{center}
		\includegraphics[width=.9\textwidth]{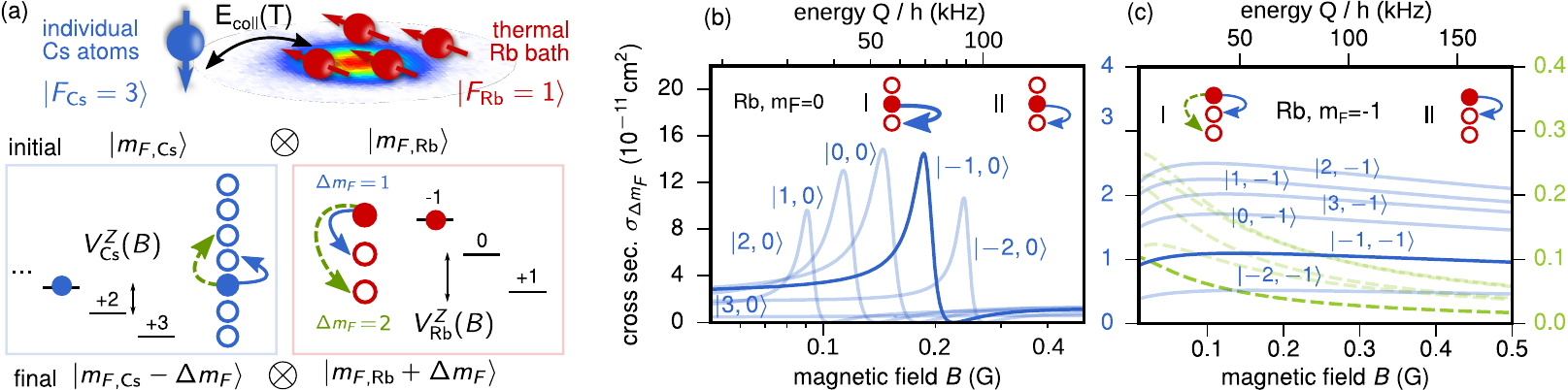}
	\end{center}
	\caption{\textbf{System overview}. 
		(a) In the experiment, individual Cs atoms collide with atoms of an ultracold, thermal Rb bath (temperature typically $\SI{450(80)}{\nano \kelvin}$, atom number $3.0(6)\times10^3$). 
		Elastic collisions between Cs and Rb lead to thermalization of Cs in Rb. 
		Additionally, in a spin-exchange (SE) collision, Cs (bottom left) and Rb (bottom right) can exchange angular momentum by $\Delta m_F=1$ (blue, solid) or $\Delta m_F=2$ (green, dashed).
		Interaction parameters, i.e. collision energy, magnetic field and internal states, are experimentally controlled, which allows tailoring the dynamics.
		In a SE collision with $\Delta \mF=1$, the energy $Q = |V_{\mathrm{Rb}}^Z - V_{\mathrm{Cs}}^Z| = h \times \SI{350}{\kilo \hertz \per G} \, \times B$ is released.
		(b)  Collision cross sections for Rb in $\mFRb=0$, where only $\Delta m_F=1$ is allowed, showing Feshbach resonances (\textit{I}) and quasi-constant SE cross sections (\textit{II}).
		(c) For Rb in $\mFRb=-1$, at low $B$ fields $\Delta m_F = +2$ processes become accessible 
		\changed{as well} (\textit{I}, dashed lines). 
	}
	\label{fig:Fig1}
\end{figure*}
Scattering channels at such small energies couple to the last bound state in the molecular potential. An unusually small binding energy below $h\times \SI {1}{\kilo \hertz}$ leads to a giant molecular wavefunction in the $\SI{}{\micro \meter}$ range.
Coupling to this state results in a series of Feshbach resonances in different collisional channels, energetically spaced by less than $h \times \SI{15}{\kilo \hertz} = k_B \times 350\,$nK and detected via SE.
Moreover, the hyperfine analogue of dipole-dipole coupling can drive SE in quanta of $2\hbar$ between Cs and Rb with measurable contribution of $\sigma_2$.

Experimentally, we prepare a dilute, thermal Rb bath of typically $3.0(6)\times10^3$ atoms at $T_{\mathrm{Rb}}=\SI{450(80)}{\nano \kelvin}$ in a desired hyperfine state and, independently, on average $6$ Cs impurities in the absolute energy ground state (see fig.~\ref{fig:Fig1}(a), details are given in Refs.~\cite{Mayer2018, Schmidt2018}).
Subsequently, Cs atoms are transported into the Rb bath by means of a species-specific optical potential \cite{Schmidt2016} and SE-driven dynamics of the Cs Zeeman state is studied by measuring the Cs Zeeman population as a function of the interaction duration $t_i$.
By repeating the experiment typically $100$ times for constant parameters, information on effective spin-dynamics is obtained.
In our strongly imbalanced mixture, Cs atoms exclusively interact with Rb atoms in one internal $\mFRb$ state, and correlations by a second collision with the same Rb atom are negligible, thus realizing particle state control for each collision event.
Additionally, the use of only few Cs atoms is crucial, because it avoids Cs-Cs interaction. 
In fact, the Cs intra-species scattering cross section exceeds the inter-species cross section by a factor of $\sim 20$, due to a \changed{broad negative}-field Feshbach resonance \cite{Lange2009}. 

Cs and Rb interact via the Hamiltonian \cite{Stoof1988}
\begin{equation}
H =E_{\mathrm{coll}} + \sum_{j=\mathrm{Cs, Rb}} \left( V^{Z}_j + V^{\mathrm{HFS}}_j \right) + \hint, 
\label{eq:interaction}
\end{equation}
with collisional energy $E_{\mathrm{coll}}$ and single-particle Zeeman and hyperfine energies  $V_j^Z$ and $V_j^{\mathrm{HFS}}$, respectively.
Finally, the interaction of both collision partners $\hint$ is determined by a molecular potential model, originating from inter-particle singlet and triplet potentials \cite{Takekoshi2012}.
In our experiment, we control \textit{all} parameters \changed{that determine the Cs-Rb dynamics, emerging from \eqref{eq:interaction}.~\sout{of the Cs-Rb interaction Hamiltonian}}, These are the temperature \changed{($E_{\mathrm{coll}}\propto T$ \methods)}, the magnetic field $B$ \changed{($V^{Z}_j \propto B$)}, and the internal Rb state \changed{($V^{\mathrm{HFS}}_{\mathrm{b}}$, $V^{\mathrm{Z}}_{\mathrm{b}}$)}.
In the $s$-wave limit at low collisional energy, the interaction $\hint$ can be effectively expressed in terms of asymptotic Cs (Rb) states, given by total angular momentum $\mathbf{\FCs}$ ($\mathbf{\FRb}$) with quantum number $\FCs$ ($\FRb$) and projections $\mFCs$ ($\mFRb$) as $\hint=\sum_{i=0, 1, 2} c_i (\mathbf{\FCs}\cdot \mathbf{\FRb})^i$ \changed{(for details, see \cite{Supplementary})}.
We conduct experiments in hyperfine ground states $\FCs=3$, $\FRb=1$, and denote collisional channels by $\ket{\mFCs, \mFRb}$.
For Cs-Rb distances of few $10\, a_0$ ($a_0 \approx \SI{0.5}{\angstrom}$ is the Bohr radius), the interaction energy $\hint$ reaches values that are of the order of $V^{\mathrm{HFS}}_j$, coupling $\mathbf{\FCs}$ and $\mathbf{\FRb}$ \cite{Takekoshi2012}.
The coupling can lead to state-changing collisions $\ket{\mFCs^{\prime}, \mFRb^{\prime}} = \ket{\mFCs-\Delta m_F, \mFRb+\Delta m_F}$, and the tensorial structure of $\hint$ allows processes with $\Delta m_F = 0, \pm 1, \pm 2$.
For $\Delta m_F=0$ processes, the internal states remain unchanged and the collision is elastic, setting the foundation for the description of interacting Bose gases \cite{Dalfovo1999} and thermalization, e.g. \cite{Hohmann2017}.
SE processes with $\Delta m_F \neq 0$ exchange angular momentum between the collision partners (see fig.~\ref{fig:Fig1}(a)), which is the basis for spinor dynamics \cite{Stamper2013}.\\
Energy and angular momentum conservation restrict observable SE processes, as shown in Fig.~\ref{fig:Fig1}~(a).
At low magnetic fields $B$, Zeeman energies $V_j^Z = \mu_j \mF_j \, B$ of Cs and Rb determine the direction of SE due to their different magnetic moments $\mu_j$ ($\mu_{\mathrm{Rb}}=2\mu_{\mathrm{Cs}}$).
SE processes with $\Delta \mF=1,2$ are exoergic, while SE with $\Delta m_F = -1, -2$ is endoergic and energetically forbidden for magnetic fields used here.
Exoergic processes $\Delta \mF = 1,2$ are further restricted by angular momentum conservation.
For Rb in $\mFRb=0$, only $\Delta \mF = 1$ is allowed, while for $\mRb=-1$, both $\Delta \mF = 1, 2$ processes are accessible.
By contrast, for $\mFRb=1$, SE with positive $\Delta \mF$ is forbidden.
Thus, the magnetic field $B$ and the choice of the Rb $\mFRb$ state grant control over collisional phenomena.
Scattering cross sections $\sigma_{\Delta m_F}$ for respective SE processes are calculated in a coupled-channel scattering model and shown in Fig.~\ref{fig:Fig1}(b),(c).
The calculations are based on a Cs-Rb interaction potential model, obtained from more than $30\times10^3$ spectroscopy lines and Feshbach resonances \cite{Takekoshi2012}.\\
For Rb bath atoms prepared in $\mFRb=0$ (see Fig.~\ref{fig:Fig1}(b)), our coupled-channel simulations indicate a regime of constant SE cross-section (regime \textit{II}, $B>\SI{300}{\milli G}$), whereas a series of Feshbach resonances in various collision channels $\ket{\mFCs, \mFRb}$ emerges at magnetic fields below $\SI{300}{\milli G}$ (regime \textit{I}).
The series originates from a coupling of the asymptotic state $\ket{\mFCs, \mFRb}$ to the first molecular bound state below the dissociation threshold, with total angular momentum $F=2$ and a binding energy as low as $-h \times \SI{490}{\hertz}$ at the magnetic field $B=\SI{80}{\milli G}$, where the first resonance occurs. 
At these low binding energies, the molecular wave function is highly delocalized with its mean radius calculated to be approximately $4000\,a_0 \approx \SI{2}{\micro \meter}$.
Note that the resonance also yields an enhanced elastic scattering cross section.
\changed{For ultracold Cs-Cs collisions, resonances have been found in a similar low-field regime \cite{Gensemer2012, Bennett2017}, which is the result of the large reduced masses in both systems (Rb-Cs and Cs-Cs). 
These lead to a dense spectrum of bound states below the dissociation threshold, which makes the occurrence of Feshbach resonances probable.}\\
For bath atoms prepared in $\mFRb=-1$ (see Fig.~\ref{fig:Fig1}(c)) SE with $\Delta m_F = 2$ arises, showing highest cross sections at magnetic fields $\leq \SI{50}{\milli G}$.
\changed{
This $\Delta m_F = 2$ SE process is so far unreported for collisions of alkali atoms.
In fact, for alkalis, scattering cross sections for dipolar spin relaxation, leading to $2\hbar$ processes in dipolar gases \cite{Hensler2003}, are at least three orders of magnitude smaller than of SE \cite{Supplementary}, thus negligible in our system.}

As the elastic cross section $\sigma_0$ exceeds SE by a factor of $\sim 10$, thermalization of Cs impurities to the Rb temperature is ensured in the presence of exoergic SE (see \methods). 
This allows modeling the time evolution of an initially prepared Cs atom in state $\mFCs=3$ with population $N_{\mFCs}$, driven by SE with Rb.
The model bases on a rate equation, where all possible SE processes $\Delta m_F$ are incorporated (see \methods).
The collision rates for elastic and SE processes $\Gamma_{\Delta m_F}$, entering this model, are directly calculated from cross sections $\Gamma_{\Delta m_F} = \sigma \left< |\mathbf{v}_{\mathrm{rel}}|\right> \nexp$, with expectation values of relative collision velocities $ \left< |\mathbf{v}_{\mathrm{rel}}|\right>$  of thermalized Cs atoms and the independently obtained Cs-Rb density overlap $\nexp$ (see \methods).\\
\begin{figure}
	\begin{center}
		\includegraphics[width=.49\textwidth]{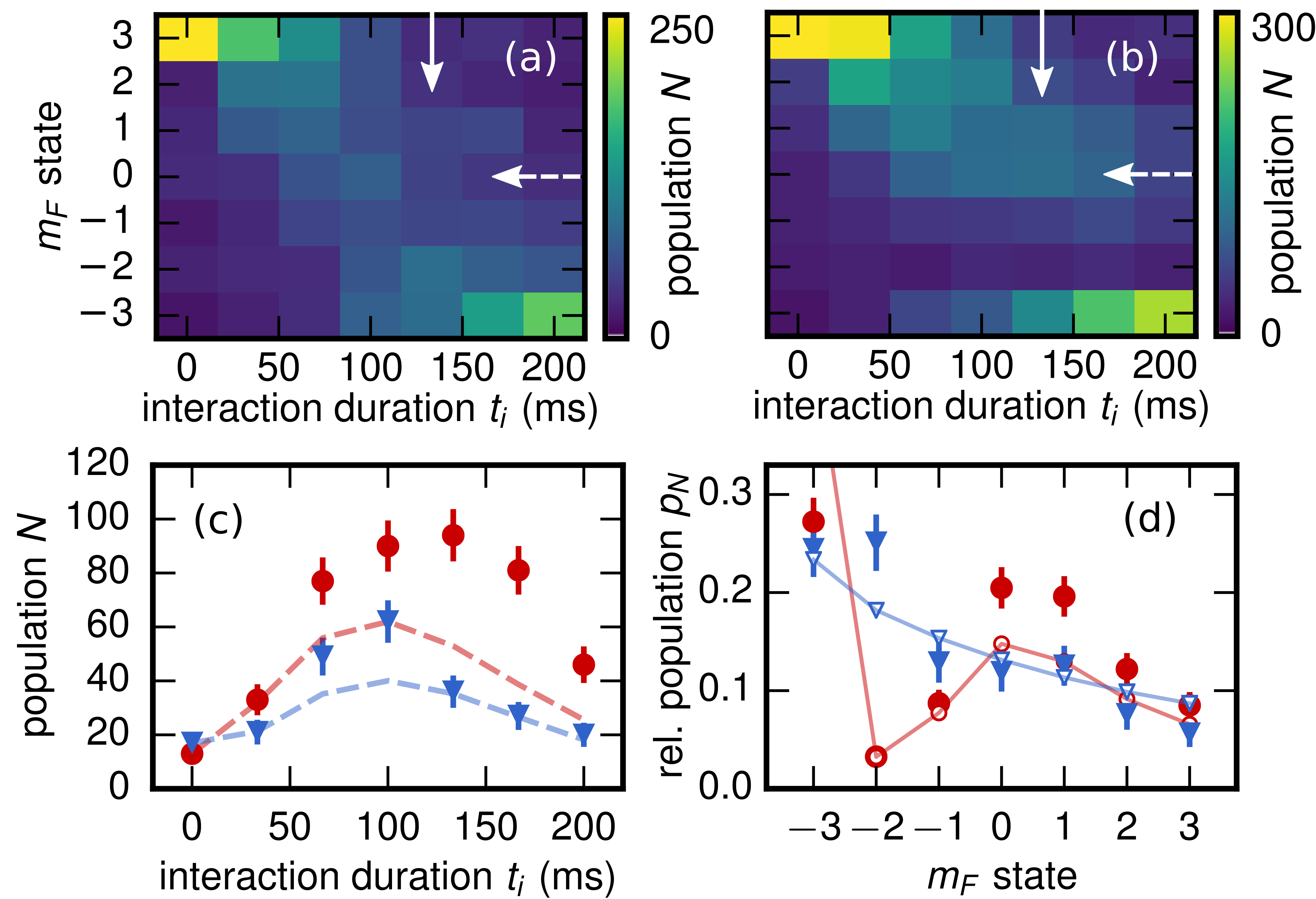}
	\end{center}
	\caption{\textbf{Spin control in the Feshbach resonance regime.} SE dynamics of Cs atoms (initially in $\mFCs=3$), immersed in a Rb bath of $\mFRb=0$. 
		The color code indicates the total measured Cs population $N$ for respective $\mFCs$ states.\\
		(a) Measurement at $B=\SI{440}{\milli G}$, showing uniform $\mFCs$ evolution. 
		SE occurs at rate $\approx \Gamma_{+1}=\SI{26}{\hertz}$ for our density overlap of $\nexp=\SI{2e12}{\centi \meter^{-3}}$.
		(b) Measurement at $B=\SI{220}{\milli G}$ shows a metastable $\mFCs=0$ state, where the population freezes in the vicinity of the zero-value of the SE scattering section of the $\ket{-1, 0}$ Feshbach resonance (compare Fig.~\ref{fig:Fig1}, (b)).
		(c) Time evolution of the $\mFCs=0$ population (dashed arrow in (a, b)) for $B$ fields of (a) (triangle) and (b) (circles) compared to results of our coupled-channel model (dashed lines, no free parameters), demonstrating the emergence of a metastable state $\mFCs=0$ \changed{in the resonance regime (b)}. 
		(d) $\mFCs$ population for a given interaction time $t_i=\SI{120}{\milli \second}$ (solid arrow in (a, b)), showing enhanced population in $\mFCs=0$ and suppressed population in $\mFCs=-2$.
		Open symbols show the same model as in (c), with solid lines guiding the eye.
		\changed{Atom counts in (a)-(c) result from the repetition of the experiment under same conditions} and error bars give statistical count uncertainties.
		}
	\label{fig:Fig2}
\end{figure}

We first explore SE phenomena for the Rb bath in $\mFRb=0$, starting at the high $B$-field regime (regime \textit{II} in fig.~\ref{fig:Fig1}(b)).
Cs atoms are prepared in the $\mFCs=3$ state initially and SE is resolved temporally for a constant magnetic field of 440\,mG (fig.~\ref{fig:Fig2}(a)).
\changed{In the absence of Feshbach resonances, SE cross sections $\sigma_{1}$ of consecutive $\ket{\mFCs, \mFRb=0}$ entrance channels are of similar magnitude.
Therefore, Cs atoms are subsequently pumped into the final $\mFCs=-3$ state in a chain of SE events.
}
As SE is unidirectional, Cs uniformly samples the full quasi-spin space $\mFCs$ and the chain of collision events is encoded in the final $\mFCs$ state.
\changed{Thus, due to the fixed ratio of elastic versus SE events, the $\mFCs$ state after the interaction also serves as an elastic-collision counter.
Interestingly, since collisional cross section are state dependent, this collision probe has non-Markovian character}.
The measured time evolution is well-reproduced by our model (see fig.~\ref{fig:Fig2}(c), (d)).

A very different picture emerges, when the spin evolution of Cs, initially in $\mFCs=3$,  is recorded at a smaller magnetic field of 220\,mG (see fig.~\ref{fig:Fig2}(b)).
Here, \changed{\sout{SE does not lead to the pumping of the $\mFCs=-3$ state, but}} the population splits into two parts, one remaining in $\mFCs=0$, while another part is pumped to $\mFCs=-3$.
This emergence of a metastable $\mFCs$ (here $\mFCs=0$) state is a direct hallmark of a magnetic Feshbach resonance in our system (regime \textit{I} in Fig.~\ref{fig:Fig1}(b)):
In the vicinity of Feshbach resonances in multiple scattering channels (compare fig.~\ref{fig:Fig1}(b)), SE cross section of the $\ket{0, 0}$ and $\ket{-1, 0}$ channels are suppressed for a wide range of collisional energies due to the zero-value of scattering cross sections $\sigma_{1}$. At the same time, interaction in the $\ket{-2, 0}$ channel is strongly enhanced at the same $B$ field, leading to a fast depopulation.
Consequently, SE does not lead to a uniform pumping \changed{to \sout{of}} the $\mFCs=-3$ state (regime \textit{I}). 
We include the thermal distribution of collision energies $p(E_{\mathrm{coll}})$ into our model (see \methods) and find excellent agreement with our measurements (see fig.~\ref{fig:Fig2}(c, d)).

\begin{figure}[b]
	\begin{center}
		\includegraphics[width=.49\textwidth]{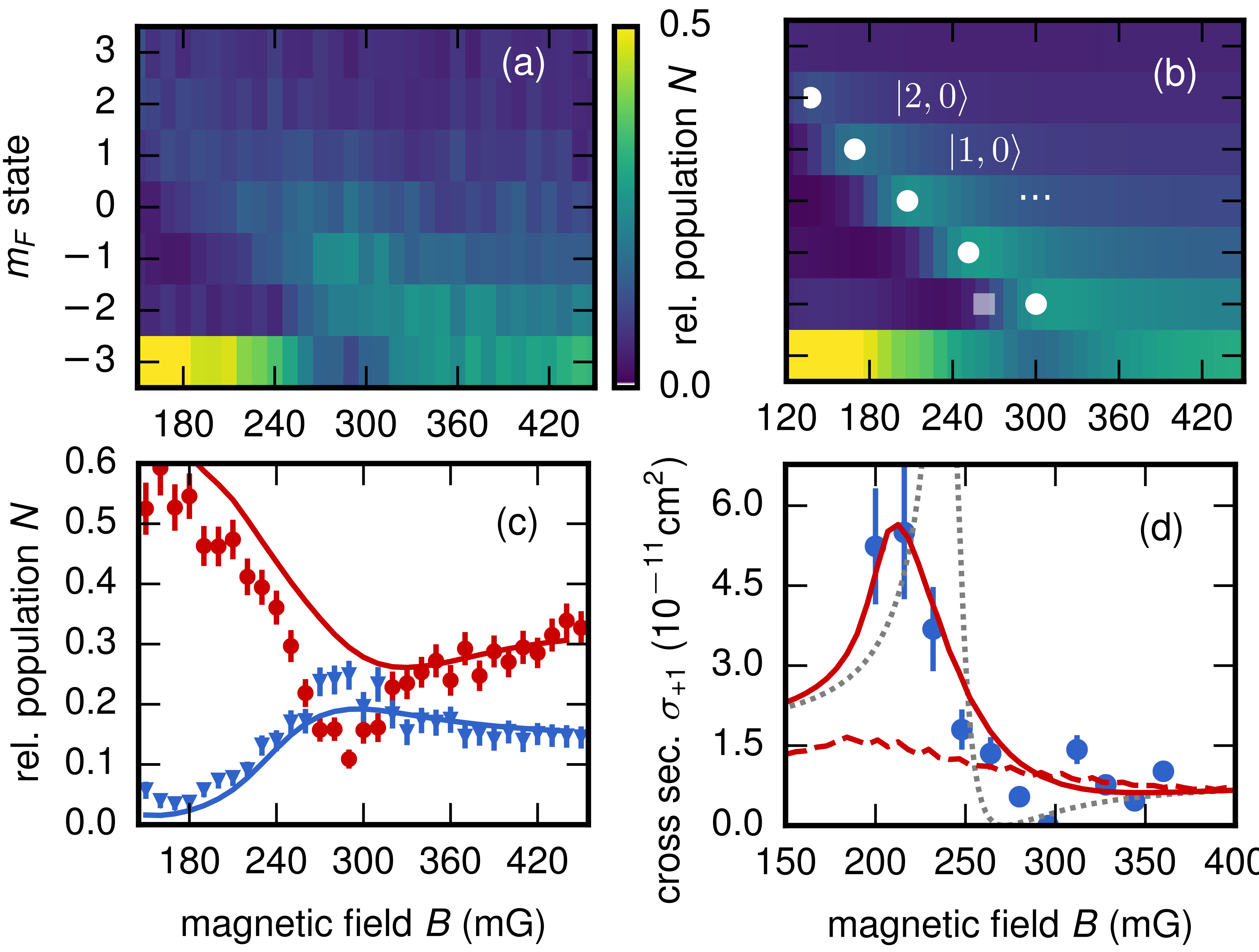}
	\end{center}
	\caption{\textbf{Low-energy Feshbach spectroscopy}. Spin-population of Cs (initially in $\mFCs=3$) immersed in a Rb $\mFRb=0$ bath for $t_i=\SI{100}{\milli \second}$ and various magnetic fields.
		(a) Metastable states, where the evolution towards the global energy minimum with $\mFCs=-3$ is interrupted in multiple $\ket{\mFCs, \mFRb=0}$ channels, indicating zero-values of respective Feshbach resonances (marked in (b)).
		High $\mF=-3$ population at low magnetic fields results from enhanced SE cross sections in all channels at low $B$ fields and thermal broadening (see fig.~\ref{fig:Fig1}(b)).
		(b) Coupled channel scattering model corresponding to measurement in (a), including thermally distributed relative collision energies (no free parameters), which shifts the Feshbach zero-value (white square) towards higher $B$ fields (white circles).
		(c) Comparison of measurement (a) (data points) and our model (b) (solid lines) for two Cs states $\mFCs=-3$ (red) and $\mFCs=-1$ (blue), showing excellent agreement.
		(d) Direct measurement of the scattering cross section for $\Delta m_F=1$ in the $\ket{-2, 0}$ channel, by measuring $\mFCs$ populations for two interaction times $t_i$ (see \methods).
		Lines show the coupled channel results for a fixed collision energy (gray), and temperature-broadened model for $T = \SI{450}{\nano \kelvin}$ (red, solid) and $T=\SI{1.5}{\micro \kelvin}$ (red, dashed).
	}
	\label{fig:Fig3}
\end{figure}
In order to find resolution limitations, we probe the energy distance between the Feshbach resonances in subsequent channels. 
Therefore, we  scan the magnetic field over the range of expected Feshbach resonances (see fig.~\ref{fig:Fig3}(a)) for a constant Cs-Rb interaction time. 
We find population enhancement in $\mFCs$ states to emerge in all scattering channels, where we expect Feshbach resonances (see Fig.~\ref{fig:Fig3}~(c)), as discussed before.
A clear distinction of individual resonances is only possible for narrowly distributed collision energies $p(E_{\mathrm{coll}})$ at ultra-low temperatures. By contrast, if the thermal spread $\Delta E_{\mathrm{coll}} \approx \sqrt{3/2} k_B T $ strongly exceeds the width of the Feshbach resonance $\gamma \Delta B$ ($\gamma$ is the magnetic moment of the Feshbach bound state, see \methods), the thermally averaged cross section does not show a minimum, thus it cannot be identified via a metastable Zeeman population.
In fact, for $T=\SI{450}{\nano \kelvin}$ ($\Delta E_{\mathrm{coll}} / h = \SI{9.4}{\kilo \hertz}$) the occupation of metastable states is less pronounced for lower-lying Feshbach resonances, where the width of the Feshbach resonances is decreasing, e.g. 27\,mG ($\SI{9.5}{\kilo \hertz}$) for $\ket{-2, 0}$ channel versus 10\,mG ($\SI{3.5}{\kilo \hertz}$) for $\ket{2, 0}$ (see Fig.~\ref{fig:Fig3}~(a)).
The influence of thermal broadening on the resolution is shown in Figure~\ref{fig:Fig3}(d). 
Here, we compare a direct measurement of the SE cross section $\sigma_1$ in the collision channel $\ket{-2, 0}$ with our finite-temperature model \changed{\cite{Supplementary}} and find excellent agreement.
By contrast, already at a bath temperature of $\SI{1.5}{\micro \kelvin}$ ($\Delta E_{\mathrm{coll}}  / h = \SI{31}{\kilo \hertz}$) thermal broadening impedes the resolution of individual Feshbach resonances, underlining the necessity to employ thermalized impurities at ultra-low temperatures.\\
\begin{figure}[tb]
	\begin{center}
		\includegraphics[width=.49\textwidth]{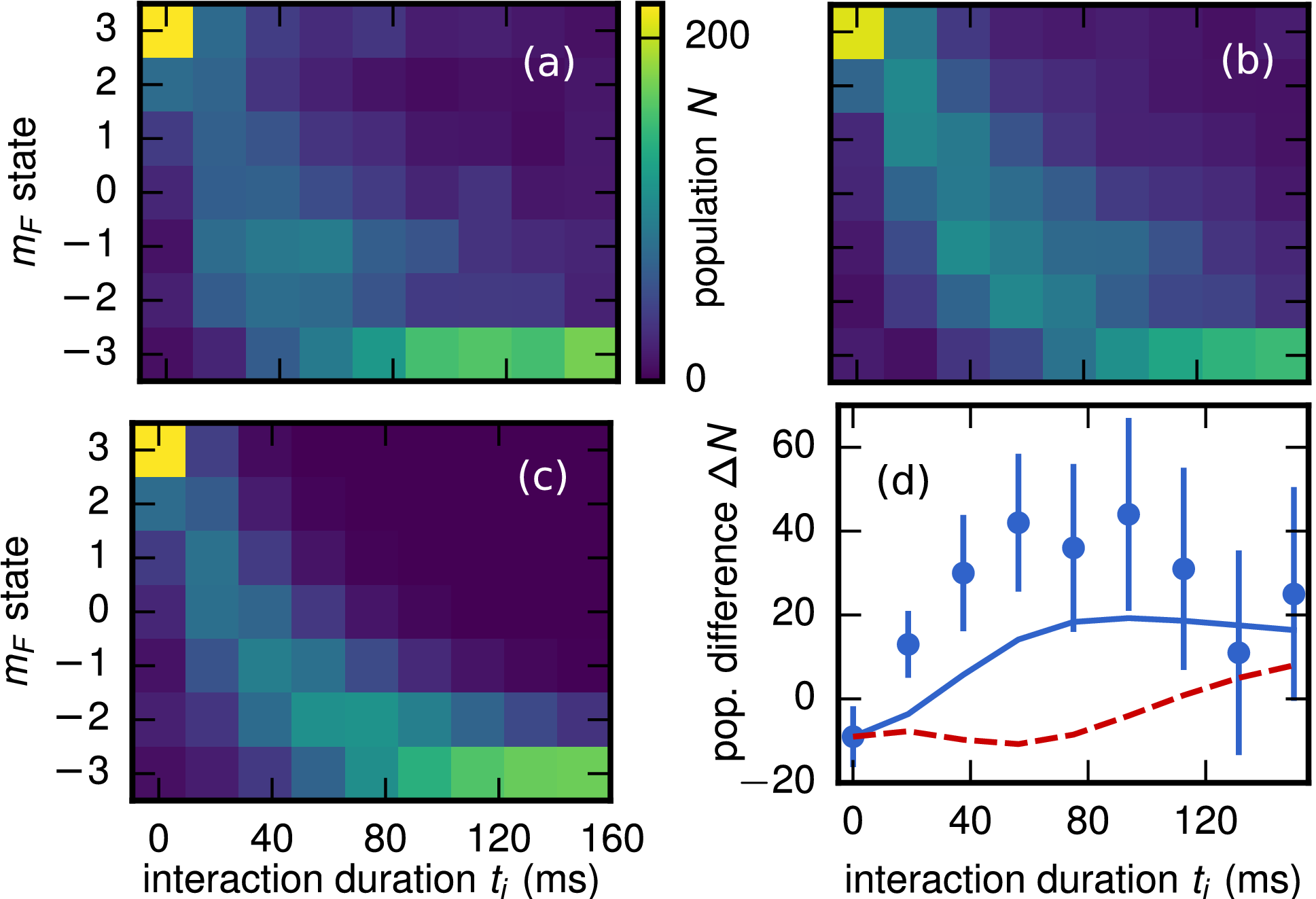}
	\end{center}
	\caption{\textbf{SE measurement with $\Delta m_F=2$}.
		Spin-evolution of Cs $\mFCs=3$ in Rb $\mFRb=-1$ bath, measured for (a) $B=\SI{50}{\milli G}$ and (b) $B=\SI{250}{\milli G}$.
		(c) Modeled evolution (no free parameters) for settings in (a).
		(d) Difference of low and high $B$ field evolution in $\mFCs=-3$ state. 
		Positive values mean faster evolution at low $B$ field values (data points). 
		Solid lines give the expectation from our model, including $2\hbar$ SE (blue, solid), and a model, where $2\hbar$ processes are excluded (red dashed), for comparison.
		\changed{Counts and errors, see Fig.~\ref{fig:Fig2}}.}
	\label{fig:Fig4}
\end{figure}
Finally, we turn to the situation of a Rb bath in $\mFRb=-1$, where SE processes in quanta of $2\hbar$ ($\Delta m_F=2$) become significant for low magnetic fields ($B < \SI{50}{\milli G}$). 
In order to distinguish $\Delta m_F=2$ SE from $\Delta m_F=1$ processes ($\sigma_1 \approx 10 \sigma_2$), two data sets are taken at magnetic fields at 50\,mG and 250\,mG (see fig.~\ref{fig:Fig4}(a), (b)), respectively. 
At these fields, SE rates for $\Delta m_F=1$ processes are the same to a level of $\approx\SI{1}{\percent}$, while $\sigma_2$ cross sections differ by a factor of two.
As a consequence, we expect a faster spin evolution for Cs at the lower magnetic field, driven by the influence of $\Delta m_F=2$ SE.
In our experiment, we realize the same bath conditions for both magnetic fields by iteratively changing the background field in the measurement throughout a total of $14\times10^3$ independent runs. 
We compare both SE series, calculating the population difference $N_{\mFCs}^{\SI{50}{\milli G}} - N_{\mFCs}^{\SI{250}{\milli G}}$.
Thereby, we effectively exclude the influence of $\Delta m_F=1$ and find a faster $\mFCs=-3$ pumping for the lower magnetic field, as expected (see fig.~\ref{fig:Fig4}(d)).
The behavior is reproduced by the full rate model, including both $\Delta m_F=1$ and $\Delta m_F=2$ processes.
By contrast, when excluding $\Delta m_F=2$ from our model, the difference in $\mFCs=-3$ pumping is negligible and the measured faster population of $\mFCs=-3$ at low magnetic field cannot be reproduced.
We conclude that our observation reveals SE processes in quanta of $2\hbar$, driven by hyperfine interaction, only.
Controlling individual impurity-bath collisions at $h \times \si{\kilo \hertz}$ energies has enabled the exploration of new SE regimes, with numerous future perspectives.
Ultralow energies allow studying and controlling individual reaction processes with Zeeman-state resolution.
Furthermore, the collision energy $E_{\mathrm{coll}}$ is tunable by accelerating Cs impurities in a species-specific transport lattice \cite{Schmidt2018}, allowing for high collision energies at ultra-low temperatures.
Finally, at even lower energies, endoergic SE with $\Delta m_F = -1$ becomes appreciable, turning the impurities' spin-state manifold into a local, highly sensitive probe of the bath's kinetic energy distribution. 
This might facilitate, for instance, the probing of quantum many-body relaxation by impurity immersion, when the bath has been driven out-of-equilibrium.\\

We thank Michael Hohmann and Axel Pelster for helpful discussions.
This work was funded in the early stage by the European Union via ERC Starting grant "QuantumProbe" and in the final stage by Deutsche Forschungsgemeinschaft via Sonderforschungsbereich (SFB) SFB/TRR185. 
D.M. and F.S. acknowledge partial funding  via  SFB/TRR49, T.L. acknowledges funding by Carl Zeiss Stiftung, and F.S. acknowledges funding by the Studienstiftung des deutschen Volkes.

\bibliography{./bibliography}{}

\begin{thebibliography}{36}%
\makeatletter
\providecommand \@ifxundefined [1]{%
 \@ifx{#1\undefined}
}%
\providecommand \@ifnum [1]{%
 \ifnum #1\expandafter \@firstoftwo
 \else \expandafter \@secondoftwo
 \fi
}%
\providecommand \@ifx [1]{%
 \ifx #1\expandafter \@firstoftwo
 \else \expandafter \@secondoftwo
 \fi
}%
\providecommand \natexlab [1]{#1}%
\providecommand \enquote  [1]{``#1''}%
\providecommand \bibnamefont  [1]{#1}%
\providecommand \bibfnamefont [1]{#1}%
\providecommand \citenamefont [1]{#1}%
\providecommand \href@noop [0]{\@secondoftwo}%
\providecommand \href [0]{\begingroup \@sanitize@url \@href}%
\providecommand \@href[1]{\@@startlink{#1}\@@href}%
\providecommand \@@href[1]{\endgroup#1\@@endlink}%
\providecommand \@sanitize@url [0]{\catcode `\\12\catcode `\$12\catcode
  `\&12\catcode `\#12\catcode `\^12\catcode `\_12\catcode `\%12\relax}%
\providecommand \@@startlink[1]{}%
\providecommand \@@endlink[0]{}%
\providecommand \url  [0]{\begingroup\@sanitize@url \@url }%
\providecommand \@url [1]{\endgroup\@href {#1}{\urlprefix }}%
\providecommand \urlprefix  [0]{URL }%
\providecommand \Eprint [0]{\href }%
\providecommand \doibase [0]{http://dx.doi.org/}%
\providecommand \selectlanguage [0]{\@gobble}%
\providecommand \bibinfo  [0]{\@secondoftwo}%
\providecommand \bibfield  [0]{\@secondoftwo}%
\providecommand \translation [1]{[#1]}%
\providecommand \BibitemOpen [0]{}%
\providecommand \bibitemStop [0]{}%
\providecommand \bibitemNoStop [0]{.\EOS\space}%
\providecommand \EOS [0]{\spacefactor3000\relax}%
\providecommand \BibitemShut  [1]{\csname bibitem#1\endcsname}%
\let\auto@bib@innerbib\@empty
\bibitem [{\citenamefont {Moses}\ \emph {et~al.}(2016)\citenamefont {Moses},
  \citenamefont {Covey}, \citenamefont {Miecnikowski}, \citenamefont {Jin},\
  and\ \citenamefont {Ye}}]{Moses2016}%
  \BibitemOpen
  \bibfield  {author} {\bibinfo {author} {\bibfnamefont {S.~A.}\ \bibnamefont
  {Moses}}, \bibinfo {author} {\bibfnamefont {J.~P.}\ \bibnamefont {Covey}},
  \bibinfo {author} {\bibfnamefont {M.~T.}\ \bibnamefont {Miecnikowski}},
  \bibinfo {author} {\bibfnamefont {D.~S.}\ \bibnamefont {Jin}}, \ and\
  \bibinfo {author} {\bibfnamefont {J.}~\bibnamefont {Ye}},\ }\href {\doibase
  10.1038/nphys3985} {\bibfield  {journal} {\bibinfo  {journal} {Nature
  Physics}\ }\textbf {\bibinfo {volume} {13}},\ \bibinfo {pages} {13} (\bibinfo
  {year} {2016})}\BibitemShut {NoStop}%
\bibitem [{\citenamefont {Bohn}\ \emph {et~al.}(2017)\citenamefont {Bohn},
  \citenamefont {Rey},\ and\ \citenamefont {Ye}}]{Bohn2017}%
  \BibitemOpen
  \bibfield  {author} {\bibinfo {author} {\bibfnamefont {J.~L.}\ \bibnamefont
  {Bohn}}, \bibinfo {author} {\bibfnamefont {A.~M.}\ \bibnamefont {Rey}}, \
  and\ \bibinfo {author} {\bibfnamefont {J.}~\bibnamefont {Ye}},\ }\href
  {\doibase 10.1126/science.aam6299} {\bibfield  {journal} {\bibinfo  {journal}
  {Science}\ }\textbf {\bibinfo {volume} {357}},\ \bibinfo {pages} {1002}
  (\bibinfo {year} {2017})}\BibitemShut {NoStop}%
\bibitem [{\citenamefont {Baranov}\ \emph {et~al.}(2012)\citenamefont
  {Baranov}, \citenamefont {Dalmonte}, \citenamefont {Pupillo},\ and\
  \citenamefont {Zoller}}]{Baranov2012}%
  \BibitemOpen
  \bibfield  {author} {\bibinfo {author} {\bibfnamefont {M.~A.}\ \bibnamefont
  {Baranov}}, \bibinfo {author} {\bibfnamefont {M.}~\bibnamefont {Dalmonte}},
  \bibinfo {author} {\bibfnamefont {G.}~\bibnamefont {Pupillo}}, \ and\
  \bibinfo {author} {\bibfnamefont {P.}~\bibnamefont {Zoller}},\ }\href
  {\doibase 10.1021/cr2003568} {\bibfield  {journal} {\bibinfo  {journal}
  {Chemical Reviews}\ }\textbf {\bibinfo {volume} {112}},\ \bibinfo {pages}
  {5012} (\bibinfo {year} {2012})}\BibitemShut {NoStop}%
\bibitem [{\citenamefont {Quemener}\ and\ \citenamefont
  {Julienne}(2012)}]{Quemener2012}%
  \BibitemOpen
  \bibfield  {author} {\bibinfo {author} {\bibfnamefont {G.}~\bibnamefont
  {Quemener}}\ and\ \bibinfo {author} {\bibfnamefont {P.~S.}\ \bibnamefont
  {Julienne}},\ }\href@noop {} {\bibfield  {journal} {\bibinfo  {journal}
  {Chemical Reviews}\ }\textbf {\bibinfo {volume} {112}},\ \bibinfo {pages}
  {4949} (\bibinfo {year} {2012})}\BibitemShut {NoStop}%
\bibitem [{\citenamefont {Safronova}\ \emph {et~al.}(2018)\citenamefont
  {Safronova}, \citenamefont {Budker}, \citenamefont {DeMille}, \citenamefont
  {Kimball}, \citenamefont {Derevianko},\ and\ \citenamefont
  {Clark}}]{Safronova2018}%
  \BibitemOpen
  \bibfield  {author} {\bibinfo {author} {\bibfnamefont {M.~S.}\ \bibnamefont
  {Safronova}}, \bibinfo {author} {\bibfnamefont {D.}~\bibnamefont {Budker}},
  \bibinfo {author} {\bibfnamefont {D.}~\bibnamefont {DeMille}}, \bibinfo
  {author} {\bibfnamefont {D.~F.~J.}\ \bibnamefont {Kimball}}, \bibinfo
  {author} {\bibfnamefont {A.}~\bibnamefont {Derevianko}}, \ and\ \bibinfo
  {author} {\bibfnamefont {C.~W.}\ \bibnamefont {Clark}},\ }\href {\doibase
  10.1103/RevModPhys.90.025008} {\bibfield  {journal} {\bibinfo  {journal}
  {Rev. Mod. Phys.}\ }\textbf {\bibinfo {volume} {90}},\ \bibinfo {pages}
  {025008} (\bibinfo {year} {2018})}\BibitemShut {NoStop}%
\bibitem [{\citenamefont {Weiner}\ \emph {et~al.}(1999)\citenamefont {Weiner},
  \citenamefont {Bagnato}, \citenamefont {Zilio},\ and\ \citenamefont
  {Julienne}}]{Weiner1999}%
  \BibitemOpen
  \bibfield  {author} {\bibinfo {author} {\bibfnamefont {J.}~\bibnamefont
  {Weiner}}, \bibinfo {author} {\bibfnamefont {V.~S.}\ \bibnamefont {Bagnato}},
  \bibinfo {author} {\bibfnamefont {S.}~\bibnamefont {Zilio}}, \ and\ \bibinfo
  {author} {\bibfnamefont {P.~S.}\ \bibnamefont {Julienne}},\ }\href {\doibase
  10.1103/RevModPhys.71.1} {\bibfield  {journal} {\bibinfo  {journal} {Rev.
  Mod. Phys.}\ }\textbf {\bibinfo {volume} {71}},\ \bibinfo {pages} {1}
  (\bibinfo {year} {1999})}\BibitemShut {NoStop}%
\bibitem [{\citenamefont {Hall}\ and\ \citenamefont
  {Willitsch}(2012)}]{Hall2012}%
  \BibitemOpen
  \bibfield  {author} {\bibinfo {author} {\bibfnamefont {F.~H.~J.}\
  \bibnamefont {Hall}}\ and\ \bibinfo {author} {\bibfnamefont {S.}~\bibnamefont
  {Willitsch}},\ }\href {\doibase 10.1103/PhysRevLett.109.233202} {\bibfield
  {journal} {\bibinfo  {journal} {Phys. Rev. Lett.}\ }\textbf {\bibinfo
  {volume} {109}},\ \bibinfo {pages} {233202} (\bibinfo {year}
  {2012})}\BibitemShut {NoStop}%
\bibitem [{\citenamefont {Wolf}\ \emph {et~al.}(2017)\citenamefont {Wolf},
  \citenamefont {Dei{\ss}}, \citenamefont {Kr{\"u}kow}, \citenamefont
  {Tiemann}, \citenamefont {Ruzic}, \citenamefont {Wang}, \citenamefont
  {D{\textquoteright}Incao}, \citenamefont {Julienne},\ and\ \citenamefont
  {Denschlag}}]{Wolf2017}%
  \BibitemOpen
  \bibfield  {author} {\bibinfo {author} {\bibfnamefont {J.}~\bibnamefont
  {Wolf}}, \bibinfo {author} {\bibfnamefont {M.}~\bibnamefont {Dei{\ss}}},
  \bibinfo {author} {\bibfnamefont {A.}~\bibnamefont {Kr{\"u}kow}}, \bibinfo
  {author} {\bibfnamefont {E.}~\bibnamefont {Tiemann}}, \bibinfo {author}
  {\bibfnamefont {B.~P.}\ \bibnamefont {Ruzic}}, \bibinfo {author}
  {\bibfnamefont {Y.}~\bibnamefont {Wang}}, \bibinfo {author} {\bibfnamefont
  {J.~P.}\ \bibnamefont {D{\textquoteright}Incao}}, \bibinfo {author}
  {\bibfnamefont {P.~S.}\ \bibnamefont {Julienne}}, \ and\ \bibinfo {author}
  {\bibfnamefont {J.~H.}\ \bibnamefont {Denschlag}},\ }\href {\doibase
  10.1126/science.aan8721} {\bibfield  {journal} {\bibinfo  {journal}
  {Science}\ }\textbf {\bibinfo {volume} {358}},\ \bibinfo {pages} {921}
  (\bibinfo {year} {2017})}\BibitemShut {NoStop}%
\bibitem [{\citenamefont {Sikorsky}\ \emph {et~al.}(2018)\citenamefont
  {Sikorsky}, \citenamefont {Meir}, \citenamefont {Ben-Shlomi}, \citenamefont
  {Akerman},\ and\ \citenamefont {Ozeri}}]{Sikorsky2018}%
  \BibitemOpen
  \bibfield  {author} {\bibinfo {author} {\bibfnamefont {T.}~\bibnamefont
  {Sikorsky}}, \bibinfo {author} {\bibfnamefont {Z.}~\bibnamefont {Meir}},
  \bibinfo {author} {\bibfnamefont {R.}~\bibnamefont {Ben-Shlomi}}, \bibinfo
  {author} {\bibfnamefont {N.}~\bibnamefont {Akerman}}, \ and\ \bibinfo
  {author} {\bibfnamefont {R.}~\bibnamefont {Ozeri}},\ }\href {\doibase
  10.1038/s41467-018-03373-y} {\bibfield  {journal} {\bibinfo  {journal}
  {Nature Communications}\ }\textbf {\bibinfo {volume} {9}},\ \bibinfo {pages}
  {920} (\bibinfo {year} {2018})}\BibitemShut {NoStop}%
\bibitem [{\citenamefont {Liu}\ \emph {et~al.}(2018)\citenamefont {Liu},
  \citenamefont {Hood}, \citenamefont {Yu}, \citenamefont {Zhang},
  \citenamefont {Hutzler}, \citenamefont {Rosenband},\ and\ \citenamefont
  {Ni}}]{Liu2018}%
  \BibitemOpen
  \bibfield  {author} {\bibinfo {author} {\bibfnamefont {L.~R.}\ \bibnamefont
  {Liu}}, \bibinfo {author} {\bibfnamefont {J.~D.}\ \bibnamefont {Hood}},
  \bibinfo {author} {\bibfnamefont {Y.}~\bibnamefont {Yu}}, \bibinfo {author}
  {\bibfnamefont {J.~T.}\ \bibnamefont {Zhang}}, \bibinfo {author}
  {\bibfnamefont {N.~R.}\ \bibnamefont {Hutzler}}, \bibinfo {author}
  {\bibfnamefont {T.}~\bibnamefont {Rosenband}}, \ and\ \bibinfo {author}
  {\bibfnamefont {K.-K.}\ \bibnamefont {Ni}},\ }\href {\doibase
  10.1126/science.aar7797} {\bibfield  {journal} {\bibinfo  {journal}
  {Science}\ }\textbf {\bibinfo {volume} {360}},\ \bibinfo {pages} {900}
  (\bibinfo {year} {2018})}\BibitemShut {NoStop}%
\bibitem [{\citenamefont {Schlagm{\"{u}}ller}\ \emph
  {et~al.}(2016)\citenamefont {Schlagm{\"{u}}ller}, \citenamefont {Liebisch},
  \citenamefont {Engel}, \citenamefont {Kleinbach}, \citenamefont
  {B{\"{o}}ttcher}, \citenamefont {Hermann}, \citenamefont {Westphal},
  \citenamefont {Gaj}, \citenamefont {L{\"{o}}w}, \citenamefont {Hofferberth},
  \citenamefont {Pfau}, \citenamefont {P{\'{e}}rez-R{\'{i}}os},\ and\
  \citenamefont {Greene}}]{Schlagmuller2016}%
  \BibitemOpen
  \bibfield  {author} {\bibinfo {author} {\bibfnamefont {M.}~\bibnamefont
  {Schlagm{\"{u}}ller}}, \bibinfo {author} {\bibfnamefont {T.~C.}\ \bibnamefont
  {Liebisch}}, \bibinfo {author} {\bibfnamefont {F.}~\bibnamefont {Engel}},
  \bibinfo {author} {\bibfnamefont {K.~S.}\ \bibnamefont {Kleinbach}}, \bibinfo
  {author} {\bibfnamefont {F.}~\bibnamefont {B{\"{o}}ttcher}}, \bibinfo
  {author} {\bibfnamefont {U.}~\bibnamefont {Hermann}}, \bibinfo {author}
  {\bibfnamefont {K.~M.}\ \bibnamefont {Westphal}}, \bibinfo {author}
  {\bibfnamefont {A.}~\bibnamefont {Gaj}}, \bibinfo {author} {\bibfnamefont
  {R.}~\bibnamefont {L{\"{o}}w}}, \bibinfo {author} {\bibfnamefont
  {S.}~\bibnamefont {Hofferberth}}, \bibinfo {author} {\bibfnamefont
  {T.}~\bibnamefont {Pfau}}, \bibinfo {author} {\bibfnamefont {J.}~\bibnamefont
  {P{\'{e}}rez-R{\'{i}}os}}, \ and\ \bibinfo {author} {\bibfnamefont {C.~H.}\
  \bibnamefont {Greene}},\ }\href {\doibase 10.1103/PhysRevX.6.031020}
  {\bibfield  {journal} {\bibinfo  {journal} {Physical Review X}\ }\textbf
  {\bibinfo {volume} {6}},\ \bibinfo {pages} {031020} (\bibinfo {year}
  {2016})}\BibitemShut {NoStop}%
\bibitem [{\citenamefont {Kleinbach}\ \emph {et~al.}(2018)\citenamefont
  {Kleinbach}, \citenamefont {Engel}, \citenamefont {Dieterle}, \citenamefont
  {L\"ow}, \citenamefont {Pfau},\ and\ \citenamefont
  {Meinert}}]{Kleinbach2018}%
  \BibitemOpen
  \bibfield  {author} {\bibinfo {author} {\bibfnamefont {K.~S.}\ \bibnamefont
  {Kleinbach}}, \bibinfo {author} {\bibfnamefont {F.}~\bibnamefont {Engel}},
  \bibinfo {author} {\bibfnamefont {T.}~\bibnamefont {Dieterle}}, \bibinfo
  {author} {\bibfnamefont {R.}~\bibnamefont {L\"ow}}, \bibinfo {author}
  {\bibfnamefont {T.}~\bibnamefont {Pfau}}, \ and\ \bibinfo {author}
  {\bibfnamefont {F.}~\bibnamefont {Meinert}},\ }\href {\doibase
  10.1103/PhysRevLett.120.193401} {\bibfield  {journal} {\bibinfo  {journal}
  {Phys. Rev. Lett.}\ }\textbf {\bibinfo {volume} {120}},\ \bibinfo {pages}
  {193401} (\bibinfo {year} {2018})}\BibitemShut {NoStop}%
\bibitem [{\citenamefont {Bennett}\ \emph {et~al.}(2017)\citenamefont
  {Bennett}, \citenamefont {Gibble}, \citenamefont {Kokkelmans},\ and\
  \citenamefont {Hutson}}]{Bennett2017}%
  \BibitemOpen
  \bibfield  {author} {\bibinfo {author} {\bibfnamefont {A.}~\bibnamefont
  {Bennett}}, \bibinfo {author} {\bibfnamefont {K.}~\bibnamefont {Gibble}},
  \bibinfo {author} {\bibfnamefont {S.}~\bibnamefont {Kokkelmans}}, \ and\
  \bibinfo {author} {\bibfnamefont {J.~M.}\ \bibnamefont {Hutson}},\ }\href
  {\doibase 10.1103/PhysRevLett.119.113401} {\bibfield  {journal} {\bibinfo
  {journal} {Phys. Rev. Lett.}\ }\textbf {\bibinfo {volume} {119}},\ \bibinfo
  {pages} {113401} (\bibinfo {year} {2017})}\BibitemShut {NoStop}%
\bibitem [{\citenamefont {Danzl}\ \emph {et~al.}(2010)\citenamefont {Danzl},
  \citenamefont {Mark}, \citenamefont {Haller}, \citenamefont {Gustavsson},
  \citenamefont {Hart}, \citenamefont {Aldegunde}, \citenamefont {Hutson},\
  and\ \citenamefont {N{\"{a}}gerl}}]{Danzl2010}%
  \BibitemOpen
  \bibfield  {author} {\bibinfo {author} {\bibfnamefont {J.~G.}\ \bibnamefont
  {Danzl}}, \bibinfo {author} {\bibfnamefont {M.~J.}\ \bibnamefont {Mark}},
  \bibinfo {author} {\bibfnamefont {E.}~\bibnamefont {Haller}}, \bibinfo
  {author} {\bibfnamefont {M.}~\bibnamefont {Gustavsson}}, \bibinfo {author}
  {\bibfnamefont {R.}~\bibnamefont {Hart}}, \bibinfo {author} {\bibfnamefont
  {J.}~\bibnamefont {Aldegunde}}, \bibinfo {author} {\bibfnamefont {J.~M.}\
  \bibnamefont {Hutson}}, \ and\ \bibinfo {author} {\bibfnamefont {H.-C.}\
  \bibnamefont {N{\"{a}}gerl}},\ }\href@noop {} {\bibfield  {journal} {\bibinfo
   {journal} {Nature Physics}\ }\textbf {\bibinfo {volume} {6}},\ \bibinfo
  {pages} {265} (\bibinfo {year} {2010})}\BibitemShut {NoStop}%
\bibitem [{\citenamefont {Ospelkaus}\ \emph {et~al.}(2010)\citenamefont
  {Ospelkaus}, \citenamefont {Ni}, \citenamefont {Wang}, \citenamefont
  {de~Miranda}, \citenamefont {Neyenhuis}, \citenamefont {Qu{\'e}m{\'e}ner},
  \citenamefont {Julienne}, \citenamefont {Bohn}, \citenamefont {Jin},\ and\
  \citenamefont {Ye}}]{Ospelkaus2010}%
  \BibitemOpen
  \bibfield  {author} {\bibinfo {author} {\bibfnamefont {S.}~\bibnamefont
  {Ospelkaus}}, \bibinfo {author} {\bibfnamefont {K.-K.}\ \bibnamefont {Ni}},
  \bibinfo {author} {\bibfnamefont {D.}~\bibnamefont {Wang}}, \bibinfo {author}
  {\bibfnamefont {M.~H.~G.}\ \bibnamefont {de~Miranda}}, \bibinfo {author}
  {\bibfnamefont {B.}~\bibnamefont {Neyenhuis}}, \bibinfo {author}
  {\bibfnamefont {G.}~\bibnamefont {Qu{\'e}m{\'e}ner}}, \bibinfo {author}
  {\bibfnamefont {P.~S.}\ \bibnamefont {Julienne}}, \bibinfo {author}
  {\bibfnamefont {J.~L.}\ \bibnamefont {Bohn}}, \bibinfo {author}
  {\bibfnamefont {D.~S.}\ \bibnamefont {Jin}}, \ and\ \bibinfo {author}
  {\bibfnamefont {J.}~\bibnamefont {Ye}},\ }\href {\doibase
  10.1126/science.1184121} {\bibfield  {journal} {\bibinfo  {journal}
  {Science}\ }\textbf {\bibinfo {volume} {327}},\ \bibinfo {pages} {853}
  (\bibinfo {year} {2010})}\BibitemShut {NoStop}%
\bibitem [{\citenamefont {Molony}\ \emph {et~al.}(2014)\citenamefont {Molony},
  \citenamefont {Gregory}, \citenamefont {Ji}, \citenamefont {Lu},
  \citenamefont {K\"oppinger}, \citenamefont {Le~Sueur}, \citenamefont
  {Blackley}, \citenamefont {Hutson},\ and\ \citenamefont
  {Cornish}}]{Molony2014}%
  \BibitemOpen
  \bibfield  {author} {\bibinfo {author} {\bibfnamefont {P.~K.}\ \bibnamefont
  {Molony}}, \bibinfo {author} {\bibfnamefont {P.~D.}\ \bibnamefont {Gregory}},
  \bibinfo {author} {\bibfnamefont {Z.}~\bibnamefont {Ji}}, \bibinfo {author}
  {\bibfnamefont {B.}~\bibnamefont {Lu}}, \bibinfo {author} {\bibfnamefont
  {M.~P.}\ \bibnamefont {K\"oppinger}}, \bibinfo {author} {\bibfnamefont
  {C.~R.}\ \bibnamefont {Le~Sueur}}, \bibinfo {author} {\bibfnamefont {C.~L.}\
  \bibnamefont {Blackley}}, \bibinfo {author} {\bibfnamefont {J.~M.}\
  \bibnamefont {Hutson}}, \ and\ \bibinfo {author} {\bibfnamefont {S.~L.}\
  \bibnamefont {Cornish}},\ }\href {\doibase 10.1103/PhysRevLett.113.255301}
  {\bibfield  {journal} {\bibinfo  {journal} {Phys. Rev. Lett.}\ }\textbf
  {\bibinfo {volume} {113}},\ \bibinfo {pages} {255301} (\bibinfo {year}
  {2014})}\BibitemShut {NoStop}%
\bibitem [{\citenamefont {Mayer}\ \emph {et~al.}(2018)\citenamefont {Mayer},
  \citenamefont {Schmidt}, \citenamefont {Adam}, \citenamefont {Haupt},
  \citenamefont {Koch}, \citenamefont {Lausch}, \citenamefont {Nettersheim},
  \citenamefont {Bouton},\ and\ \citenamefont {Widera}}]{Mayer2018}%
  \BibitemOpen
  \bibfield  {author} {\bibinfo {author} {\bibfnamefont {D.}~\bibnamefont
  {Mayer}}, \bibinfo {author} {\bibfnamefont {F.}~\bibnamefont {Schmidt}},
  \bibinfo {author} {\bibfnamefont {D.}~\bibnamefont {Adam}}, \bibinfo {author}
  {\bibfnamefont {S.}~\bibnamefont {Haupt}}, \bibinfo {author} {\bibfnamefont
  {J.}~\bibnamefont {Koch}}, \bibinfo {author} {\bibfnamefont {T.}~\bibnamefont
  {Lausch}}, \bibinfo {author} {\bibfnamefont {J.}~\bibnamefont {Nettersheim}},
  \bibinfo {author} {\bibfnamefont {Q.}~\bibnamefont {Bouton}}, \ and\ \bibinfo
  {author} {\bibfnamefont {A.}~\bibnamefont {Widera}},\ }\href@noop {}
  {\bibfield  {journal} {\bibinfo  {journal} {arXiv preprint arXiv:1805.01313}\
  } (\bibinfo {year} {2018})}\BibitemShut {NoStop}%
\bibitem [{\citenamefont {Schmidt}\ \emph {et~al.}(2018)\citenamefont
  {Schmidt}, \citenamefont {Mayer}, \citenamefont {Bouton}, \citenamefont
  {Adam}, \citenamefont {Lausch}, \citenamefont {Spethmann},\ and\
  \citenamefont {Widera}}]{Schmidt2018}%
  \BibitemOpen
  \bibfield  {author} {\bibinfo {author} {\bibfnamefont {F.}~\bibnamefont
  {Schmidt}}, \bibinfo {author} {\bibfnamefont {D.}~\bibnamefont {Mayer}},
  \bibinfo {author} {\bibfnamefont {Q.}~\bibnamefont {Bouton}}, \bibinfo
  {author} {\bibfnamefont {D.}~\bibnamefont {Adam}}, \bibinfo {author}
  {\bibfnamefont {T.}~\bibnamefont {Lausch}}, \bibinfo {author} {\bibfnamefont
  {N.}~\bibnamefont {Spethmann}}, \ and\ \bibinfo {author} {\bibfnamefont
  {A.}~\bibnamefont {Widera}},\ }\href {\doibase
  10.1103/PhysRevLett.121.130403} {\bibfield  {journal} {\bibinfo  {journal}
  {Phys. Rev. Lett.}\ }\textbf {\bibinfo {volume} {121}},\ \bibinfo {pages}
  {130403} (\bibinfo {year} {2018})}\BibitemShut {NoStop}%
\bibitem [{\citenamefont {Schmidt}\ \emph {et~al.}(2016)\citenamefont
  {Schmidt}, \citenamefont {Mayer}, \citenamefont {Hohmann}, \citenamefont
  {Lausch}, \citenamefont {Kindermann},\ and\ \citenamefont
  {Widera}}]{Schmidt2016}%
  \BibitemOpen
  \bibfield  {author} {\bibinfo {author} {\bibfnamefont {F.}~\bibnamefont
  {Schmidt}}, \bibinfo {author} {\bibfnamefont {D.}~\bibnamefont {Mayer}},
  \bibinfo {author} {\bibfnamefont {M.}~\bibnamefont {Hohmann}}, \bibinfo
  {author} {\bibfnamefont {T.}~\bibnamefont {Lausch}}, \bibinfo {author}
  {\bibfnamefont {F.}~\bibnamefont {Kindermann}}, \ and\ \bibinfo {author}
  {\bibfnamefont {A.}~\bibnamefont {Widera}},\ }\href {\doibase
  10.1103/PhysRevA.93.022507} {\bibfield  {journal} {\bibinfo  {journal} {Phys.
  Rev. A}\ }\textbf {\bibinfo {volume} {93}},\ \bibinfo {pages} {022507}
  (\bibinfo {year} {2016})}\BibitemShut {NoStop}%
\bibitem [{\citenamefont {Lange}\ \emph {et~al.}(2009)\citenamefont {Lange},
  \citenamefont {Pilch}, \citenamefont {Prantner}, \citenamefont {Ferlaino},
  \citenamefont {Engeser}, \citenamefont {N\"agerl}, \citenamefont {Grimm},\
  and\ \citenamefont {Chin}}]{Lange2009}%
  \BibitemOpen
  \bibfield  {author} {\bibinfo {author} {\bibfnamefont {A.~D.}\ \bibnamefont
  {Lange}}, \bibinfo {author} {\bibfnamefont {K.}~\bibnamefont {Pilch}},
  \bibinfo {author} {\bibfnamefont {A.}~\bibnamefont {Prantner}}, \bibinfo
  {author} {\bibfnamefont {F.}~\bibnamefont {Ferlaino}}, \bibinfo {author}
  {\bibfnamefont {B.}~\bibnamefont {Engeser}}, \bibinfo {author} {\bibfnamefont
  {H.-C.}\ \bibnamefont {N\"agerl}}, \bibinfo {author} {\bibfnamefont
  {R.}~\bibnamefont {Grimm}}, \ and\ \bibinfo {author} {\bibfnamefont
  {C.}~\bibnamefont {Chin}},\ }\href {\doibase 10.1103/PhysRevA.79.013622}
  {\bibfield  {journal} {\bibinfo  {journal} {Phys. Rev. A}\ }\textbf {\bibinfo
  {volume} {79}},\ \bibinfo {pages} {013622} (\bibinfo {year}
  {2009})}\BibitemShut {NoStop}%
\bibitem [{\citenamefont {Stoof}\ \emph {et~al.}(1988)\citenamefont {Stoof},
  \citenamefont {Koelman},\ and\ \citenamefont {Verhaar}}]{Stoof1988}%
  \BibitemOpen
  \bibfield  {author} {\bibinfo {author} {\bibfnamefont {H.~T.~C.}\
  \bibnamefont {Stoof}}, \bibinfo {author} {\bibfnamefont {J.~M. V.~A.}\
  \bibnamefont {Koelman}}, \ and\ \bibinfo {author} {\bibfnamefont {B.~J.}\
  \bibnamefont {Verhaar}},\ }\href {\doibase 10.1103/PhysRevB.38.4688}
  {\bibfield  {journal} {\bibinfo  {journal} {Phys. Rev. B}\ }\textbf {\bibinfo
  {volume} {38}},\ \bibinfo {pages} {4688} (\bibinfo {year}
  {1988})}\BibitemShut {NoStop}%
\bibitem [{\citenamefont {Takekoshi}\ \emph {et~al.}(2012)\citenamefont
  {Takekoshi}, \citenamefont {Debatin}, \citenamefont {Rameshan}, \citenamefont
  {Ferlaino}, \citenamefont {Grimm}, \citenamefont {N\"agerl}, \citenamefont
  {Le~Sueur}, \citenamefont {Hutson}, \citenamefont {Julienne}, \citenamefont
  {Kotochigova},\ and\ \citenamefont {Tiemann}}]{Takekoshi2012}%
  \BibitemOpen
  \bibfield  {author} {\bibinfo {author} {\bibfnamefont {T.}~\bibnamefont
  {Takekoshi}}, \bibinfo {author} {\bibfnamefont {M.}~\bibnamefont {Debatin}},
  \bibinfo {author} {\bibfnamefont {R.}~\bibnamefont {Rameshan}}, \bibinfo
  {author} {\bibfnamefont {F.}~\bibnamefont {Ferlaino}}, \bibinfo {author}
  {\bibfnamefont {R.}~\bibnamefont {Grimm}}, \bibinfo {author} {\bibfnamefont
  {H.-C.}\ \bibnamefont {N\"agerl}}, \bibinfo {author} {\bibfnamefont {C.~R.}\
  \bibnamefont {Le~Sueur}}, \bibinfo {author} {\bibfnamefont {J.~M.}\
  \bibnamefont {Hutson}}, \bibinfo {author} {\bibfnamefont {P.~S.}\
  \bibnamefont {Julienne}}, \bibinfo {author} {\bibfnamefont {S.}~\bibnamefont
  {Kotochigova}}, \ and\ \bibinfo {author} {\bibfnamefont {E.}~\bibnamefont
  {Tiemann}},\ }\href {\doibase 10.1103/PhysRevA.85.032506} {\bibfield
  {journal} {\bibinfo  {journal} {Phys. Rev. A}\ }\textbf {\bibinfo {volume}
  {85}},\ \bibinfo {pages} {032506} (\bibinfo {year} {2012})}\BibitemShut
  {NoStop}%
\bibitem [{Sup()}]{Supplementary}%
  \BibitemOpen
  \href@noop {} {\bibinfo  {journal} {See Supplemental Material for details on
  experimental and data analysis procedures, which includes refs. \cite{Ho1998,
  Ohmi1998, Mudrich2002, Hutson2008, Chin2010, Blackley2014, Cannoni2014,
  Tiesinga1993}}\ }\BibitemShut {NoStop}%
\bibitem [{\citenamefont {Dalfovo}\ \emph {et~al.}(1999)\citenamefont
  {Dalfovo}, \citenamefont {Giorgini}, \citenamefont {Pitaevskii},\ and\
  \citenamefont {Stringari}}]{Dalfovo1999}%
  \BibitemOpen
\bibfield  {journal} {  }\bibfield  {author} {\bibinfo {author} {\bibfnamefont
  {F.}~\bibnamefont {Dalfovo}}, \bibinfo {author} {\bibfnamefont
  {S.}~\bibnamefont {Giorgini}}, \bibinfo {author} {\bibfnamefont {L.~P.}\
  \bibnamefont {Pitaevskii}}, \ and\ \bibinfo {author} {\bibfnamefont
  {S.}~\bibnamefont {Stringari}},\ }\href {\doibase 10.1103/RevModPhys.71.463}
  {\bibfield  {journal} {\bibinfo  {journal} {Rev. Mod. Phys.}\ }\textbf
  {\bibinfo {volume} {71}},\ \bibinfo {pages} {463} (\bibinfo {year}
  {1999})}\BibitemShut {NoStop}%
\bibitem [{\citenamefont {Hohmann}\ \emph {et~al.}(2017)\citenamefont
  {Hohmann}, \citenamefont {Kindermann}, \citenamefont {Lausch}, \citenamefont
  {Mayer}, \citenamefont {Schmidt}, \citenamefont {Lutz},\ and\ \citenamefont
  {Widera}}]{Hohmann2017}%
  \BibitemOpen
  \bibfield  {author} {\bibinfo {author} {\bibfnamefont {M.}~\bibnamefont
  {Hohmann}}, \bibinfo {author} {\bibfnamefont {F.}~\bibnamefont {Kindermann}},
  \bibinfo {author} {\bibfnamefont {T.}~\bibnamefont {Lausch}}, \bibinfo
  {author} {\bibfnamefont {D.}~\bibnamefont {Mayer}}, \bibinfo {author}
  {\bibfnamefont {F.}~\bibnamefont {Schmidt}}, \bibinfo {author} {\bibfnamefont
  {E.}~\bibnamefont {Lutz}}, \ and\ \bibinfo {author} {\bibfnamefont
  {A.}~\bibnamefont {Widera}},\ }\href {\doibase
  10.1103/PhysRevLett.118.263401} {\bibfield  {journal} {\bibinfo  {journal}
  {Phys. Rev. Lett.}\ }\textbf {\bibinfo {volume} {118}},\ \bibinfo {pages}
  {263401} (\bibinfo {year} {2017})}\BibitemShut {NoStop}%
\bibitem [{\citenamefont {Stamper-Kurn}\ and\ \citenamefont
  {Ueda}(2013)}]{Stamper2013}%
  \BibitemOpen
  \bibfield  {author} {\bibinfo {author} {\bibfnamefont {D.~M.}\ \bibnamefont
  {Stamper-Kurn}}\ and\ \bibinfo {author} {\bibfnamefont {M.}~\bibnamefont
  {Ueda}},\ }\href {\doibase 10.1103/RevModPhys.85.1191} {\bibfield  {journal}
  {\bibinfo  {journal} {Rev. Mod. Phys.}\ }\textbf {\bibinfo {volume} {85}},\
  \bibinfo {pages} {1191} (\bibinfo {year} {2013})}\BibitemShut {NoStop}%
\bibitem [{\citenamefont {Gensemer}\ \emph {et~al.}(2012)\citenamefont
  {Gensemer}, \citenamefont {Martin-Wells}, \citenamefont {Bennett},\ and\
  \citenamefont {Gibble}}]{Gensemer2012}%
  \BibitemOpen
  \bibfield  {author} {\bibinfo {author} {\bibfnamefont {S.~D.}\ \bibnamefont
  {Gensemer}}, \bibinfo {author} {\bibfnamefont {R.~B.}\ \bibnamefont
  {Martin-Wells}}, \bibinfo {author} {\bibfnamefont {A.~W.}\ \bibnamefont
  {Bennett}}, \ and\ \bibinfo {author} {\bibfnamefont {K.}~\bibnamefont
  {Gibble}},\ }\href {\doibase 10.1103/PhysRevLett.109.263201} {\bibfield
  {journal} {\bibinfo  {journal} {Phys. Rev. Lett.}\ }\textbf {\bibinfo
  {volume} {109}},\ \bibinfo {pages} {263201} (\bibinfo {year}
  {2012})}\BibitemShut {NoStop}%
\bibitem [{\citenamefont {Hensler}\ \emph {et~al.}(2003)\citenamefont
  {Hensler}, \citenamefont {Werner}, \citenamefont {Griesmaier}, \citenamefont
  {Schmidt}, \citenamefont {G{\"{o}}rlitz}, \citenamefont {Pfau}, \citenamefont
  {Giovanazzi},\ and\ \citenamefont {Rzazewski}}]{Hensler2003}%
  \BibitemOpen
  \bibfield  {author} {\bibinfo {author} {\bibfnamefont {S.}~\bibnamefont
  {Hensler}}, \bibinfo {author} {\bibfnamefont {J.}~\bibnamefont {Werner}},
  \bibinfo {author} {\bibfnamefont {A.}~\bibnamefont {Griesmaier}}, \bibinfo
  {author} {\bibfnamefont {P.~O.}\ \bibnamefont {Schmidt}}, \bibinfo {author}
  {\bibfnamefont {A.}~\bibnamefont {G{\"{o}}rlitz}}, \bibinfo {author}
  {\bibfnamefont {T.}~\bibnamefont {Pfau}}, \bibinfo {author} {\bibfnamefont
  {S.}~\bibnamefont {Giovanazzi}}, \ and\ \bibinfo {author} {\bibfnamefont
  {K.}~\bibnamefont {Rzazewski}},\ }\href {\doibase 10.1007/s00340-003-1334-0}
  {\bibfield  {journal} {\bibinfo  {journal} {Applied Physics B}\ }\textbf
  {\bibinfo {volume} {77}},\ \bibinfo {pages} {765} (\bibinfo {year}
  {2003})}\BibitemShut {NoStop}%
\bibitem [{\citenamefont {Ho}(1998)}]{Ho1998}%
  \BibitemOpen
  \bibfield  {author} {\bibinfo {author} {\bibfnamefont {T.-L.}\ \bibnamefont
  {Ho}},\ }\href {\doibase 10.1103/PhysRevLett.81.742} {\bibfield  {journal}
  {\bibinfo  {journal} {Phys. Rev. Lett.}\ }\textbf {\bibinfo {volume} {81}},\
  \bibinfo {pages} {742} (\bibinfo {year} {1998})}\BibitemShut {NoStop}%
\bibitem [{\citenamefont {Ohmi}\ and\ \citenamefont
  {Machida}(1998)}]{Ohmi1998}%
  \BibitemOpen
  \bibfield  {author} {\bibinfo {author} {\bibfnamefont {T.}~\bibnamefont
  {Ohmi}}\ and\ \bibinfo {author} {\bibfnamefont {K.}~\bibnamefont {Machida}},\
  }\href {\doibase 10.1143/JPSJ.67.1822} {\bibfield  {journal} {\bibinfo
  {journal} {Journal of the Physical Society of Japan}\ }\textbf {\bibinfo
  {volume} {67}},\ \bibinfo {pages} {1822} (\bibinfo {year}
  {1998})}\BibitemShut {NoStop}%
\bibitem [{\citenamefont {Mudrich}\ \emph {et~al.}(2002)\citenamefont
  {Mudrich}, \citenamefont {Kraft}, \citenamefont {Singer}, \citenamefont
  {Grimm}, \citenamefont {Mosk},\ and\ \citenamefont
  {Weidem\"uller}}]{Mudrich2002}%
  \BibitemOpen
  \bibfield  {author} {\bibinfo {author} {\bibfnamefont {M.}~\bibnamefont
  {Mudrich}}, \bibinfo {author} {\bibfnamefont {S.}~\bibnamefont {Kraft}},
  \bibinfo {author} {\bibfnamefont {K.}~\bibnamefont {Singer}}, \bibinfo
  {author} {\bibfnamefont {R.}~\bibnamefont {Grimm}}, \bibinfo {author}
  {\bibfnamefont {A.}~\bibnamefont {Mosk}}, \ and\ \bibinfo {author}
  {\bibfnamefont {M.}~\bibnamefont {Weidem\"uller}},\ }\href {\doibase
  10.1103/PhysRevLett.88.253001} {\bibfield  {journal} {\bibinfo  {journal}
  {Phys. Rev. Lett.}\ }\textbf {\bibinfo {volume} {88}},\ \bibinfo {pages}
  {253001} (\bibinfo {year} {2002})}\BibitemShut {NoStop}%
\bibitem [{\citenamefont {Hutson}\ \emph {et~al.}(2008)\citenamefont {Hutson},
  \citenamefont {Tiesinga},\ and\ \citenamefont {Julienne}}]{Hutson2008}%
  \BibitemOpen
  \bibfield  {author} {\bibinfo {author} {\bibfnamefont {J.~M.}\ \bibnamefont
  {Hutson}}, \bibinfo {author} {\bibfnamefont {E.}~\bibnamefont {Tiesinga}}, \
  and\ \bibinfo {author} {\bibfnamefont {P.~S.}\ \bibnamefont {Julienne}},\
  }\href {\doibase 10.1103/PhysRevA.78.052703} {\bibfield  {journal} {\bibinfo
  {journal} {Phys. Rev. A}\ }\textbf {\bibinfo {volume} {78}},\ \bibinfo
  {pages} {052703} (\bibinfo {year} {2008})}\BibitemShut {NoStop}%
\bibitem [{\citenamefont {Chin}\ \emph {et~al.}(2010)\citenamefont {Chin},
  \citenamefont {Grimm}, \citenamefont {Julienne},\ and\ \citenamefont
  {Tiesinga}}]{Chin2010}%
  \BibitemOpen
  \bibfield  {author} {\bibinfo {author} {\bibfnamefont {C.}~\bibnamefont
  {Chin}}, \bibinfo {author} {\bibfnamefont {R.}~\bibnamefont {Grimm}},
  \bibinfo {author} {\bibfnamefont {P.}~\bibnamefont {Julienne}}, \ and\
  \bibinfo {author} {\bibfnamefont {E.}~\bibnamefont {Tiesinga}},\ }\href
  {\doibase 10.1103/RevModPhys.82.1225} {\bibfield  {journal} {\bibinfo
  {journal} {Rev. Mod. Phys.}\ }\textbf {\bibinfo {volume} {82}},\ \bibinfo
  {pages} {1225} (\bibinfo {year} {2010})}\BibitemShut {NoStop}%
\bibitem [{\citenamefont {Blackley}\ \emph {et~al.}(2014)\citenamefont
  {Blackley}, \citenamefont {Julienne},\ and\ \citenamefont
  {Hutson}}]{Blackley2014}%
  \BibitemOpen
  \bibfield  {author} {\bibinfo {author} {\bibfnamefont {C.~L.}\ \bibnamefont
  {Blackley}}, \bibinfo {author} {\bibfnamefont {P.~S.}\ \bibnamefont
  {Julienne}}, \ and\ \bibinfo {author} {\bibfnamefont {J.~M.}\ \bibnamefont
  {Hutson}},\ }\href {\doibase 10.1103/PhysRevA.89.042701} {\bibfield
  {journal} {\bibinfo  {journal} {Phys. Rev. A}\ }\textbf {\bibinfo {volume}
  {89}},\ \bibinfo {pages} {042701} (\bibinfo {year} {2014})}\BibitemShut
  {NoStop}%
\bibitem [{\citenamefont {Cannoni}(2014)}]{Cannoni2014}%
  \BibitemOpen
  \bibfield  {author} {\bibinfo {author} {\bibfnamefont {M.}~\bibnamefont
  {Cannoni}},\ }\href {\doibase 10.1103/PhysRevD.89.103533} {\bibfield
  {journal} {\bibinfo  {journal} {Phys. Rev. D}\ }\textbf {\bibinfo {volume}
  {89}},\ \bibinfo {pages} {103533} (\bibinfo {year} {2014})}\BibitemShut
  {NoStop}%
\bibitem [{\citenamefont {Tiesinga}\ \emph {et~al.}(1993)\citenamefont
  {Tiesinga}, \citenamefont {Verhaar},\ and\ \citenamefont
  {Stoof}}]{Tiesinga1993}%
  \BibitemOpen
  \bibfield  {author} {\bibinfo {author} {\bibfnamefont {E.}~\bibnamefont
  {Tiesinga}}, \bibinfo {author} {\bibfnamefont {B.~J.}\ \bibnamefont
  {Verhaar}}, \ and\ \bibinfo {author} {\bibfnamefont {H.~T.~C.}\ \bibnamefont
  {Stoof}},\ }\href {\doibase 10.1103/PhysRevA.47.4114} {\bibfield  {journal}
  {\bibinfo  {journal} {Phys. Rev. A}\ }\textbf {\bibinfo {volume} {47}},\
  \bibinfo {pages} {4114} (\bibinfo {year} {1993})}\BibitemShut {NoStop}%
\end{thebibliography}%

\section{Supplementary Material}
\subsection{Experimental procedure}
We prepare a thermal Rb sample of typically $3 \times 10^3$ atoms and a temperature of 450 nK in a cigar-shaped crossed dipole trap with long axis along $z$ in internal state $\ket{\FRb=1, \mFRb=0}$.
Subsequently, individual Cs atoms are trapped in a high-gradient magneto-optical trap, transferred into an independent crossed dipole trap, and pumped into the $\mFCs=3$ hyperfine ground state by means of degenerate Raman sideband cooling.
Rb is transferred into a desired Zeeman $\mFRb=0, \pm1$ state by microwave-driven Landau-Zener transitions.
Cs impurities are loaded into a species-selective, one-dimensional optical conveyor belt lattice \cite{Schmidt2016} and transported into the Rb cloud.
We expect the first elastic Cs collision within 3\,ms after switching off the transport lattice, which initiates the thermalization of the Cs atoms to the bath temperature and defines the starting point of the Cs-Rb interaction.
The lattice is extinguished after transport, leaving Cs and Rb mobile in the Rb crossed dipole trap with trap frequencies for Cs (Rb) of $\omega_z =2\pi \times \SI{53}{\hertz}, \, \omega_r =2\pi \times \SI{670}{\hertz}$ ($\omega_z =2\pi \times \SI{56}{\hertz}, \, \omega_r =2\pi \times \SI{700}{\hertz}$) in axial and radial direction, respectively.
The magnetic field $B$ direction and amplitude during the Cs-Rb interaction is controlled with an accuracy of roughly 5 mG, associated with a Zeeman energy for Cs of $\approx\SI{2}{\kilo \hertz}$. 
In addition, we ensure adiabatic ramping of magnetic fields to prevent mixing of Zeeman states.
%
We interrupt the interaction after a duration $t_i$ by removing Rb from the trap by a laser pulse of $\SI{0.5}{\milli \second}$, resonant to the Rb $D_2$ line.
After this Rb pushout the one-dimensional lattice is switched on, fixing the Cs position along the $z$ direction.
Finally, individual Cs atoms in $\mFCs$ are read out by a combination of $\mFCs$ state-sensitive microwave transitions and fluorescence imaging in the species-selective lattice.

\subsection{Details on interaction Hamiltonian}
The atomic interaction $\Hint$ comprises two contributions
$\Hint = V^c(r) + V^d$, a central Coulomb potential $V^c(r)$ and a dipolar interaction $V^d$.
\changed{$V^d$ includes magnetic dipole-dipole interaction of the valence electrons, and second order spin-orbit coupling \cite{Takekoshi2012}.}

Angular momentum is conserved in the central potential $V^c(r)$, thus the total angular momentum $\mathbf{F} = \mathbf{\FCs} + \mathbf{\FRb}$. 
The total spin quantum number $F$ and its projection to the quantization axis $M$ are good quantum numbers to label quantum states.
Therefore, the central interaction potential $V^c$ for ultracold temperatures (s-wave regime) can be expressed in terms of projections into the coupled spin basis $F$ \cite{Ho1998, Ohmi1998}, ranging from $\left| \FCs - \FRb \right|$ to $\FCs + \FRb$. 
With density overlap $\nexp$, the central potential rewrites
\newcommand{\sumF}{\sum_{F=\left| \FCs - \FRb \right|}^{\FCs + \FRb}}
$V^c = \nexp \sumF g_F \mathcal{P}_F$ with projection operators into total $F$ channels $\mathcal{P}_F = \sum_{M=-F}^{F} \ket{F, M}\bra{F, M}$ \changed{and} $g_F = \frac{4 \pi \hbar^2}{\mu} a_F$, with s-wave scattering length $a_F$ for total $F$).

\changed{
\subsection{Interaction in atomic bases}
Alternatively, instead of using eigenstates of the total angular momentum $\mathbf{F}$, i.e. $\ket{F, M}$, the interaction potential can be expressed in terms of the atomic angular momentum operators, $\mathbf{\FRb}$ and $\mathbf{\FCs}$.
The transformation is based on the relation $(\mathbf{\FRb} \cdot \mathbf{\FCs}) \times \mathbf{1}= (1/2)(\mathbf{F}^2 - \mathbf{\FRb} ^2 - \mathbf{\FCs}^2) \times \sum_{F=|\FCs-\FRb|}^{\FCs+\FRb}\mathcal{P}_F$, where $\mathbf{1}$ is the identity operator.
Exploiting the orthogonality of $\ket{F, M}$ states, we can express multiples of the $\mathbf{\FRb} \cdot \mathbf{\FCs}$ product as 
\begin{equation}
	(\mathbf{\FRb} \cdot \mathbf{\FCs})^i = \sum_{F=|\FCs-\FRb|}^{\FCs+\FRb} \lambda_F^i \mathcal{P}_F,
\end{equation}
with $\lambda_F = (1/2)(F(F+1) - \FCs(\FCs+1) - \FRb(\FRb+1))$.
This allows a mapping between projection operators $\mathcal{P}_F$ and $\mathbf{\FCs}, \, \mathbf{\FRb}$ operators via a matrix $\mathbf{V}$, with dimension $n \times n$ and $n = (\FCs + \FRb - |\FCs - \FRb| + 1)$.
For ground state collisions with $\FCs=3,\, \FRb=1$, the mapping writes
\begin{equation*}
	\left(
	\begin{matrix}
	(\mathbf{\FCs} \cdot \mathbf{\FRb})^2 \\ ... \\ \mathbf{1}
	\end{matrix}
	\right)
	= \left(
	\begin{matrix}9 & 1 & 16\\3 & -1 & -4\\1 & 1 & 1\end{matrix}
	\right)
	\times 	
	\left(
	\begin{matrix}
	\PF{F = 4} \\ ... \\  \PF{{F = 2}}
	\end{matrix}
	\right).
\end{equation*}
Based on this relation, the central $s$-wave interaction potential is expressed in terms $\mathbf{\FCs}$, $\mathbf{\FRb}$ as
\begin{equation}
V^c / \nexp= c_0 \mathbb{1} + c_{1} \mathbf{\FCs} \cdot \mathbf{\FRb} + c_2 ( \mathbf{\FCs} \cdot \mathbf{\FRb})^2,
\end{equation}
with coefficients $c_0 =-\frac{1}{7}g_2 + g_3 + \frac{1}{7} g_4$, $c_1=-\frac{2}{21}g_2 - \frac{1}{12}g_3 + \frac{5}{28} g_4$, and $c_2 =\frac{1}{21}g_2 - \frac{1}{12} g_3 + \frac{1}{28} g_4$.
The formalism used here has been shown previously in \cite{Ho1998,Ohmi1998}.}

In this form, interaction can be clearly separated into an elastic (mean-field) contribution $ c_0 \mathbb{1}$ and SE, driven by hyperfine coupling $(\mathbf{\FCs} \cdot \mathbf{\FRb} )^i$.
Therefore, it is commonly used in the description of spinor-dynamics, typically for indistinguishable bosons \cite{Stamper2013}.
The total two-particle Hamiltonian (including Zeeman and hyperfine energy) has a tensorial (\changed{rank 2}) structure, thus allowing for SE \changed{from entrance channels $\ket{\mFCs, \mFRb}$ to $|\mFCs-\Delta \mF, \mFRb+\Delta \mF \rangle$}, with $\Delta \mF=0, \pm1, \pm2$.

\subsection{Coupled channel calculations}
\FigSone
\FigStwo
Microscopic scattering cross sections $\sigma_{\Delta \mF}$ for elastic ($\Delta \mF=0$), and SE collisions  ($\Delta \mF\neq0$) have been obtained, using coupled channel scattering calculations, based on the refined model of Cs-Rb interaction \cite{Takekoshi2012}.
\changed{
In order to quantify the influence of $V^d$ on the spin dynamics via dipolar relaxation, we have initially included partial waves until $l=2$ ($d$-waves) in the coupled channel scattering calculations.
We found that cross sections for dipolar relaxation (due to $V^d$) are at least three orders of magnitude smaller than spin exchange (due to the interplay of $V^c$ and the hyperfine energy).
Thus, we have confirmed the negligible influence of dipole-dipole coupling for alkali atoms \cite{Tiesinga1993} in our system.
}
For a constant collision energy for a Rb bath temperature of 450\,nK, scattering cross sections $\sigma_{\Delta \mF}$ for allowed processes are shown in figure~\ref{fig:RatesNumeric_mF0} for Rb in $\mFRb=0$ and in figure~\ref{fig:RatesNumeric_mFm1} for Rb in $\mFRb=-1$.
Properties of weakly bound levels for Feshbach molecules are calculated using a propagation method similar to \cite{Hutson2008}.
The series of Feshbach resonances for Rb in $\mFRb=0$ originates with a closed collision channel, dominated by the least bound state ($n=-1$) with total angular momentum $\ket{F=2, M}$, where $M=\mFCs+\mFRb$ corresponds to the asymptotic channel.
At $B=\SI{80}{\milli G}$, where the first Feshbach resonance occurs for $M=2$, the binding energy with respect to the $\ket{\FCs=3, \mFCs=2} + \ket{\FRb=1, \mFRb=0}$ asymptote is $-h \times \SI{490}{\hertz}$.
In the basis of asymptotic states, $F=2$ comprises main contributions from three states, which are $\ket{\FCs=3, \mFCs=3} + \ket{\FRb=1, \mFRb=-1}$, $\ket{\FCs=3, \mFCs=2} + \ket{\FRb=1, \mFRb=0}$, and $\ket{\FCs=3, \mFCs=1} + \ket{\FRb=1, \mFRb=1}$.
Figure~\ref{fig:feshbachStates} shows the respective radial wavefunction $R_i$, weighted by their respective contribution to the $\ket{F=2, M=2}$ state. 
$\ket{F=2, M=2}$ is pre-dissociated, shown by the oscillatory behavior at large $r$ of channel $\ket{\FCs=3, \mFCs=1} + \ket{\FRb=1, \mFRb=1}$ (black, dotted curve in Fig.~\ref{fig:feshbachStates}).
The size of the $\ket{F=2, M=2}$ state is calculated from the sum of radial expectation values $\left< r_i \right>=\int_{0}^{\infty} r |R_i|^2 \diff r$.
Thus, we obtain the radius of the bound state of $\left< r \right> = \sum_{i} \left< r_i \right> = 4026\, a_0$.
\FigSthree
\end{figure}

\subsection{Thermal broadening of a Feshbach resonance}
Feshbach resonances, shown in Fig.~1 of the main text, are calculated for a fixed relative collision energy $E_{\mathrm{coll}} = \frac{1}{2} \mu |\mathbf{v}_{\mathrm{rel}}|^2$ of Cs and Rb with reduced mass $\mu$. 
At the Feshbach resonance, the collision energy $E_{\mathrm{coll}}$ is in resonance with a bound molecular state at magnetic field $B_0$.
A change of the collision energy by $\delta E_{\mathrm{coll}}$  with respect to $E_{\mathrm{coll}}$ reflects in a change of the resonance position $B_0(E) = B_0 +  \delta E_{\mathrm{coll}} /\gamma$.
The sensitivity $\gamma$ is associated with the $g$ factor of the molecular bound state and is determined from simulations at various collision energies in the following.
The dependency on the collision energy shifts the Feshbach resonance.
We use the standard model of the s-wave scattering length across the resonance of \cite{Chin2010}
\begin{equation}
a(B, E_{\mathrm{coll}}) = a_0 \left( 1 - \frac{\Delta}{B - (B_0 + \delta E_{\mathrm{coll}} /\gamma)} \right) .
\end{equation}
%
From the scattering length, the collisional cross section $\sigma$ is derived. 
With relative collision wave vector $k = \sqrt{2 \mu E_{\mathrm{coll}}} / \hbar$ the cross section writes \cite{Blackley2014}
\begin{equation}
\sigma(a, k) = \frac{8 \pi a^2}{ k^2 a^2 + \left( \frac{1}{2} k^2 r_e a - 1 \right)^2 },
\label{eq:crossSection}
\end{equation}
where $r_e$ is the effective range of the Rb-Cs van der Waals potential \cite{Blackley2014}
\begin{eqnarray}
r_e & = & \frac{[\Gamma(1/4)]^4}{6 \pi^2} \bar{a}
\left[ 1-2\frac{\bar{a}}{a_0} + 2 \left( \frac{\bar{a}}{a_0} \right)^2 \right] \\
\bar{a} & = & \frac{2\pi}{[\Gamma(1/4)]^2} \left( \frac{2 \mu C_6}{\hbar^2} \right)^{1/4},
\end{eqnarray}
derived from the van der Waals $C_6$ coefficient, taken from \cite{Takekoshi2012}.
In order to determine $\gamma$, we perform scattering calculations for three fixed collision energies $E_{\mathrm{coll}} / k_B$ of $\SI{250}{\nano \kelvin}$, $\SI{450}{\nano \kelvin}$, and $\SI{1.5}{\micro \kelvin}$.
For each setting, we fit Feshbach resonance positions $B_0$ and resonance widths $\Delta$ (see tabs.~\ref{tab:250nK}, \ref{tab:450nK}, \ref{tab:1500nK}), using eq.~\ref{eq:crossSection}, and find $\gamma = h\times \SI{350.5}{\kilo \hertz \per G}$.\\

Additionally, we include the effect of the finite temperature-induced broadening of the Feshbach resonance:
In a thermalized Cs-Rb mixture at temperature $T$, collision energies $E_{\mathrm{coll}}$ are statistically distributed, following a Maxwell-Boltzmann distribution \cite{Cannoni2014} $p(E_{\mathrm{coll}}) \propto \sqrt{E} \exp{(-E_{\mathrm{coll}} / k_B T)}$, with expectation value $\left< E_{\mathrm{coll}} \right> = (3/2) \, k_B T$ and spread (standard deviation) $\Delta E_{\mathrm{coll}} = ({\left< E_{\mathrm{coll}}^2 \right> - \left< E_{\mathrm{coll}} \right>^2})^{1/2} = \sqrt{3/2} k_B T$. 
We eliminate the energy dependency of collision cross section $\sigma_{\Delta \mF}(B, E_{\mathrm{coll}})$ by calculating the expectation value $\sigma(B) = \int  p(E_{\mathrm{coll}}) \sigma_{\Delta \mF}(B, E_{\mathrm{coll}}) dE_{\mathrm{coll}}$.

\subsection{Spin-exchange model}
Our experimental observable is the internal Cs impurity state $\mFCs$, driven by SE with Rb.
We model the population $N_{\mFCs}$ evolution in each of the 7 available Zeeman sub states in the $\FCs=3$ hyperfine manifold.
The evolution $\dot{N}_{\mFCs}$ is driven by SE between adjacent states.
In general, processes with $\Delta m_F = \pm1, \pm2 $ are possible, which we include into a rate model
\begin{equation}
\begin{aligned}
	\dot{N}_{\mFCs} = 
	- &\sum_{\Delta \mFCs = \pm 1, \pm 2} \spinrate_{\mFCs, m_{F, \mathrm{Rb}}}^{\mFCs - \Delta \mf, m_{F, \mathrm{Rb}} + \Delta \mf} N_{\mFCs} \\
	+ &\sum_{\Delta \mf = \pm 1, \pm 2} \spinrate_{\mFCs + \Delta \mf, m_{F, \mathrm{Rb}}}^{\mFCs, m_{F, \mathrm{Rb}}  + \Delta \mf} N_{\mFCs + \Delta \mf}   \\
	- &\Lambda N_{\mFCs}.
	\label{eq:RateModel}
\end{aligned}
\end{equation}
Rates $\spinrate_{m_{F, \mathrm{Cs}}, m_{F, \mathrm{Rb}}}^{m_{\mathrm{F, Cs}}^{\prime}, m_{\mathrm{F, Rb}}^{\prime}}$ give the absolute values of SE rates for the transition $\ket{\mFCs, \mFRb} \rightarrow \ket{\mFCs^{\prime}, \mFRb^{\prime}}$.
Positive and negative sign ($+, -$) denote population gain and loss, respectively.
Integrating eq.~\ref{eq:RateModel} over time yields time-dependent $\mFCs$ populations without any free parameters, used as a model in the main text.

Three-body recombination of one Cs atoms with two Rb atoms is included by the loss rate $\Lambda$.
Note, that endoergic processes $\Delta \mF=-1, -2$ are forbidden for magnetic fields used here ($B>\SI{27}{\milli G}$) at average collision energies $E_{\mathrm{coll}}$.
Rates for elastic and SE collisions $\Gamma_{\Delta \mF=0, 1, 2}$ are related to microscopic scattering cross sections $\sigma_{{\Delta \mF}}$ via $\Gamma_{\Delta \mF} = \left< \sigma_{{\Delta \mF}}  |\mathbf{v}_{\mathrm{rel}}| \densrb \right>$.
Here, $\left< \cdot \right>$ denotes the expectation value of the scattering rate with the relative velocity of Cs and Rb atoms $|\mathbf{v}_{\mathrm{rel}}|$ and local Rb density $\densrb$.
In thermal baths, the velocity and density distribution of atoms are not correlated and the scattering cross section is a constant and rates write $\Gamma_{\Delta \mF=0, 1, 2} = \sigma_{\Delta \mF} \left<|\mathbf{v}_{\mathrm{rel}}|\right> \left< n \right>$.
For thermal Cs and Rb atoms at temperature $\TCs$ and $\TRb$, respectively, the mean thermal velocity writes \cite{Mudrich2002} 
\begin{equation}
	\left< |\mathbf{v}_{\mathrm{rel}}| \right> = \sqrt{{8 \kB}/{\pi} \left({\TCs}/{\mRb} + {\TRb}/{\mCs} \right)}
	\label{eq:velocityExpVal}
\end{equation} with Cs mass $\mCs$ and Rb mass $\mRb$. 
The density-density overlap of Cs and Rb at density $\denscs$ and $\densrb$, respectively, writes $\nexp = \int \denscs \densrb d^{3}\mathbf{r}$
and is calculated, assuming thermal, Gaussian distributed clouds in a harmonic trap.
In our model, we assume fully thermalized Cs atoms at Rb bath temperature, see discussion below.

\subsection{Direct measurement of a resonance}
In figure~3~(d) of the manuscript, we show the result of a direct measurement of the SE scattering cross section around a Feshbach resonance, going beyond the evidence of resonances via the detection of metastable $\mFCs$ states (result of zero-values of SE cross sections, see fig.~3 of main text).
We focus on the $\spinrate_{-2, 0}^{-3, 1}$ ($\ket{\mFCs=-2, \mFRb=0} \rightarrow \ket{\mFCs=-3, \mFRb=1}$) resonance.
Expressing the rate in terms of the microscopic exchange constant $\spinrate_{-2, 0}^{-3, 1} = 
\sigma_{-2, 0}^{-3, 1} \left<v\right> \left<n\right>$, the time evolution of the $\mFCs = -3$ population is given by a reduced rate model 
\begin{equation}
	\dot{N}_{-3} = - \lossrate N_{-3}  + \spinrate_{-2, m_2}^{-3, m_2+1} N_{-2}.
\end{equation}
Note, that the $\Delta \mF = 2$ rate constant is zero for Rb in $\mFRb = 0$.
We extract the rate constant by measuring the population in $N_{-3}$ and $N_{-2}$ for two interaction durations $t_1$, $t_2$.
The evolution time $\Delta t = t_2 - t_1$ is chosen to yield an average in the order of one spin-exchange event for the given $\Delta t$.
We extract the exchange cross section from $\sigma_{-2, 0}^{-3, 1} = \frac{\dot{N}_{-3}}{\left< v\right>\left< n \right> N_{-2}} \approx \frac{N_{-3}(t_2) - N_{-3}(t_1)}{\Delta t \left< n \right> N_{-2}(t_1)}$.
Error bars in figure~3~(d) are extracted via standard maximal error propagation.

\subsection{Influence of exoergic collisions on Cs velocity distribution}
In our atomic mixture, spin-exchange releases the energy $Q$ into the system, accelerating Cs and Rb after a collision, while elastic collisions lead to an equilibration of momenta. 
For an ensemble of Cs impurities, this counter play should yield a steady equilibrium state, which is of relevance for precision measurements of the SE rates, e.g. shown in figure 4.
In order to include effects of thermalization (elastic) and driving (SE), we perform a Monte-Carlo simulation, where trajectories of independent Cs atoms in the (harmonically approximated) optical dipole trap are integrated in time, driven by elastic and SE collisions.
The simulation includes elastic collisions at rate $\Gammael$ and spin-exchange at rate $\Gammase$, calculated from scattering cross sections, shown in figure~{\ref{fig:RatesNumeric_mF0}}.
In each simulation time step $\Delta t$ the probability for an elastic (spin-exchange collision) is evaluated from the probability $P_{\mathrm{el}} = 1 - \mathrm{exp}(-\Gammael \Delta t)$ ($P_{\mathrm{se}} = 1 - \mathrm{exp}(-\Gammase \Delta t)$).
In the center-of-mass system, the collision leads to an isotropic redistribution of momentum in a collision.
The isotropy is modeled by randomly choosing the direction of momenta of both collision partners.
After the collision, the momenta of Cs (Rb) are $\mathbf{p}_{\mathrm{Cs}}^{\prime}$ ($\mathbf{p}_{\mathrm{Rb}}^{\prime}$).
In the case of spin-exchange, the absolute values of the center-of-mass momenta increase due to the super-elastic character of the collision according to
\begin{equation}
\frac{|\mathbf{p}_{\mathrm{Cs}}^{\prime}|^2}{2 \mu} = \frac{|\mathbf{p}_{\mathrm{Cs}}^{\,}|^2}{2 \mu} + Q,
\end{equation}
where $\mathbf{p}_{\mathrm{Cs}}^{\,}$ is the initial momentum in the center-of-mass frame, $\mu$ is the reduced mass and $Q$ the energy released during the spin-exchange of $Q$. 
We simulate $10^4$ independent Cs trajectories in a Rb bath with homogeneous density $\densrb$ and temperature $\TRb=\SI{300}{\nano \kelvin}$ for different magnetic fields.
The value $\frac{\pi m}{8 k_B}\left< v \right>^2$ is a measure for the average kinetic energy of the particles. 
For Maxwell-Boltzmann distributed Cs velocities $v$, it corresponds to the temperature $\TCs$ of the cloud.
We compare measurements with magnetic fields of 50\,mG and 250\,mG, as used in the measurement in Fig.~4 and study the time evolution of the Cs velocity distribution.
Results are given in Figure~\ref{fig:thermalization} and show that the energy of the  Cs ensemble equilibrates after a short interaction duration $t_i$.
Note, that the timescale to reach an equilibrium state of about $2-4 \, \si{\milli \second}$ is in the order of the inverse rate of spin-exchange $\tau_{\mathrm{se}}=\SI{3.1}{\milli \second}$ for the Rb bath density chosen here.
We compare the Cs velocity distribution with a Maxwell-Boltzmann distribution, assuming the average kinetic energy as $\TCs$, which reproduces the distribution.
The increase of Cs temperature in the presence of SE collisions is in the order of $\SI{17}{\percent}$ above the Rb temperature of $\SI{450}{\nano \kelvin}$, which yields a change of the relative collision velocity $|\mathbf{v}_{\mathrm{rel}}|$ in equation~\ref{eq:velocityExpVal} of $<\SI{4}{\percent}$, and we neglect the effect in our model.
In fact, for the spin-evolution shown in Fig.~4, the temperature increase would yield faster SE for the higher $B$ field, which cannot explain our observation of faster SE at lower $B$ field.
\FigSfour

\subsection{Feshbach resonance parameters}
Feshbach resonance positions $B_0$, widths $\Delta$ and elastic background scattering length $a$ have been fitted from cross sections in figure~\ref{fig:RatesNumeric_mF0}~(a).
$\sigma_{B_0, \Delta, a}$ give the fitting uncertainty of the respective values.
\begin{table}[h]
	\caption{\textbf{Feshbach resonance}, $E_{\mathrm{coll}} = k_B  \times \SI{250}{\nano \kelvin}$}
	\label{tab:250nK}
	\begin{center}
		\begin{tabular}{|c | ccccc |}
			\hline
			$(\mFCs, \mFRb)$ & (-2, 0) & (-1, 0)& (0, 0)&(1, 0)&(2, 0) \\
			\hline
			$B_0$ (mG)            &225.66 &  173.00 &  131.62 &  100.73 &  78.40 \\
			$\sigma_{B_0}$ (mG)   &0.00 &    0.01 &    0.01 &    0.01 &   0.00 \\
			$\Delta$  (mG)        &25.18 &   29.11 &   27.14 &   20.14 &  10.32 \\
			$\sigma_{\Delta}$ (mG)&0.01 &    0.02 &    0.02 &    0.03 &   0.01 \\
			$a$ ($a_0$)           &968.51 & 1050.81 & 1056.32 & 1010.84 & 950.16 \\
			$\sigma_{a}$ ($a_0$)  &0.25 &    0.47 &    0.41 &    0.73 &   0.41 \\
			\hline
		\end{tabular}
	\end{center}
\end{table}
\begin{table}[h]
	\caption{\textbf{Feshbach resonance}, $E_{\mathrm{coll}} = k_B  \times \SI{450}{\nano \kelvin}$}
	\label{tab:450nK}
	\begin{center}
		\begin{tabular}{|c | ccccc |}
			\hline
			$(\mFCs, \mFRb)$ & (-2, 0) & (-1, 0)& (0, 0)&(1, 0)&(2, 0) \\
			\hline
			$B_0$ (mG)              &235.80 & 184.91 &   141.59 & 112.68 &  90.29 \\
			$\sigma_{B_0}$ (mG)    &0.00 &   0.01 &     0.01 &   0.01 &   0.01 \\
			$\Delta$  (mG)          &26.90 &  29.04 &    28.90 &  19.91 &  10.24 \\
			$\sigma_{\Delta}$ (mG) &0.01 &   0.01 &     0.03 &   0.03 &   0.02 \\
			$a$ ($a_0$)          &-994.43 & 992.89 & -1074.50 & 968.12 & 924.27 \\
			$\sigma_{a}$ ($a_0$) &0.13 &   0.28 &     0.62 &   0.78 &   0.83 \\
			\hline
		\end{tabular}
	\end{center}
\end{table}
\begin{table}[h]
	\caption{\textbf{Feshbach resonance}, $E_{\mathrm{coll}} = k_B  \times \SI{1.5}{\micro \kelvin}$}
	\label{tab:1500nK}
	\begin{center}
		\begin{tabular}{|c | ccccc |}
			\hline
			$(\mFCs, \mFRb)$ & (-2, 0) & (-1, 0)& (0, 0)&(1, 0)&(2, 0) \\
			\hline
			$B_0$ (mG)              &299.39 & 247.37 & 206.42 & 175.36 & 153.04 \\
			$\sigma_{B_0}$ (mG)   &0.11 &   0.17 &   0.20 &   0.15 &   0.18 \\
			$\Delta$  (mG)            &24.56 &  27.75 &  25.37 &  18.66 &   9.54 \\
			$\sigma_{\Delta}$ (mG)&0.19 &   0.28 &   0.32 &   0.29 &   0.18 \\
			$a$ ($a_0$)          &872.80 & 907.25 & 916.92 & 904.41 & 876.84 \\
			$\sigma_{a}$ ($a_0$) &3.79 &   5.39 &   6.77 &   8.36 &   8.91 \\
			\hline
		\end{tabular}
	\end{center}
\end{table}

\newpage
\bibliographystyle{apsrev4-1}

\end{document}